\documentclass[showpacs,twocolumn,preprintnumbers,aps,superscriptaddress]{revtex4}
\usepackage{graphicx}
\usepackage{epsfig}
\usepackage{color}
\usepackage{hyperref}
\usepackage{graphicx}
\usepackage{amsmath}
\usepackage{amssymb}
\usepackage{amsfonts}

\graphicspath{{./fig-jpg/}{./fig-ps/}{./figures/}{./Figs/}}

\DeclareMathOperator{\sech}{sech}
\definecolor{mygreen}{rgb}{0,0.6,0}

\begin{document}

\title{Dynamics and stabilization of bright soliton stripes in the
hyperbolic-dispersion nonlinear Schr\"{o}dinger equation}
\date{\today}
\author{L. A. Cisneros-Ake}
\affiliation{Departamento de Matem\'aticas, ESFM, Instituto Polit\'{e}cnico Nacional,
Unidad Profesional Adolfo L\'{o}pez Mateos Edificio 9, 07738 Cd.~de
M\'exico, M\'{e}xico}
\email{cisneros@esfm.ipn.mx}
\author{R. Carretero-Gonz{\'a}lez}
\affiliation{Nonlinear Dynamical Systems Group,\thanks{\texttt{URL}: http://nlds.sdsu.edu}
Computational Sciences Research Center, and Department of Mathematics and
Statistics, San Diego State University, San Diego, California 92182-7720, USA}
\author{P. G. Kevrekidis}
\affiliation{Department of Mathematics and Statistics, University of Massachusetts,
Amherst, Massachusetts 01003-4515, USA}
\author{B. A. Malomed}
\affiliation{Department of Physical Electronics, Faculty of Engineering, and Center for
Light-Matter Interaction, Tel Aviv University, Tel Aviv 69978, Israel }

\begin{abstract}
We consider the dynamics and stability of bright soliton stripes in the
two-dimensional nonlinear Schr\"{o}dinger equation with hyperbolic
dispersion, under the action of transverse perturbations. We start by
discussing a recently proposed
adiabatic-invariant approximation for transverse instabilities and
its limitations in the bright soliton case. We then focus on
a variational approximation used to
reduce the dynamics of the bright-soliton stripe to effective equations of
motion for its transverse shift. The reduction allows us to address the
stripe's snaking instability, which is inherently present in the system, and
follow the ensuing spatiotemporal undulation dynamics. Further, introducing
a channel-shaped potential, we show that the instabilities (not only flexural,
but also those of the necking type) can be attenuated, up to the point of
complete stabilization of the soliton stripe.
\end{abstract}

\maketitle

\section{Introduction}

Nonlinear Schr\"{o}dinger (NLS) equations are a class of universal models
governing the nonlinear propagation of waves in dispersive and diffractive
media~\cite{sulem,ablowitz}.
In particular, in one dimension (1D) the interplay of the
self-focusing cubic nonlinearity and paraxial diffraction, or,
alternatively, anomalous group-velocity dispersion (GVD) gives rise to the
commonly known bright solitons~\cite{Zakharov}, which are replaced by dark
solitons in the case of the normal GVD~\cite{Agrawal,siam}. The NLS equations
play an equally important role in 2D and 3D settings, where the diffraction
and GVD act together, in the combination with the cubic self-focusing or
defocusing~\cite{review,sulem,JiankeBook:2010,Fibich,siam}. In the case of the anomalous sign of the GVD and
self-focusing sign of the nonlinearity, 2D and 3D solitons (in particular,
``light bullets'' in optics~\cite{bullets}) exist too as
formal solutions to the multidimensional NLS equations, but they are
strongly unstable. 

An alternative possibility is to consider the interplay of the paraxial
diffraction in one transverse direction (assuming that the setting is
two-dimensional) with the normal GVD and self-focusing, which is usually
called, as a whole, the NLS equation with the \textit{hyperbolic} 2D
dispersion.
The latter is a model of wide
interest~\cite{sulem,CoDiTri,CoTri,GHS1,GHS2,GHS3}, appearing in a number
of different applications. These include but are not limited to
deep water waves~\cite{abseg,zakh}, cyclotron
waves in plasmas~\cite{sen,myra} and nonlinear optics~\cite{trillo1,trillo2},
to name a few.
The hyperbolicity is realized as the opposite signs of the
diffraction and GVD terms in the equation, see Eq.~(\ref{eq:NLS}) below. In
this case, 2D bright solitons do not exist, but one can consider quasi-1D
solitons, which are self-trapped in the transverse direction, and uniformly
extended along the longitudinal coordinate. Because such 
\textit{bright-soliton stripes} 
may have obviously important physical realizations,
a significant problem concerns the stability of the stripes~\cite{Kuznetsov:1986}. It
is well known that they are actually subject to the \textit{snaking}
(flexural) instability.

An initial purpose of the present work is to extend the known analytical
approach to the description of the long-wave snaking instability, so as to
explore its manifestations for perturbations with finite wavelengths, as
well as the development of the instability beyond the usual limit of the
linear approximation. The main objective of the work is to elaborate a
possibility to attenuate and, eventually, completely suppress the
instability by trapping the stripe in an underlying channel potential.
This is somewhat in the spirit of our earlier effort to stabilize dark solitons
via a potential barrier in the self-defocusing NLS case~\cite{dark1},
although the stabilization is trickier here, as we will explain in detail
below, due to the accessibility of the solitary waves to different
types of instabilities.

The rest of the paper is organized as follows.
The model, its basic features and theoretical setup
are given in Sec.~\ref{sec:theory}.
Effective longitudinal
equations of motion for the soliton stripes are derived in 
Sec.~\ref{sec:reduced}, using the method based on the use 
of the adiabatic invariant, recently quite successfully used
in the case of dark solitons. 
Subsequently, a more elaborate
(and more accurate in the present setting) variational approximation
is developed. In Sec.~\ref{sec:numerics}
we numerically investigate the dynamics and stability, as produced by
different effective stripe-evolution equations, and compare the results with
those directly produced by the underlying NLS equation. 
In Sec.~\ref{sec:potential} 
we introduce an appropriately crafted external potential
that is able to control (and eventually eliminate) all potential
instabilities. Finally, in Sec.~\ref{sec:conclu} we give a brief recap of
the results and outline directions for further work.

\section{Model and Theoretical Setup}
\label{sec:theory}

As per the above discussion, our
model of choice will be the 2D NLS equation with the hyperbolic dispersion in the
presence of the external potential, $V(x,y)$:
\begin{equation}
iu_{t}=-\frac{1}{2}u_{xx}+\frac{1}{2}u_{yy}-|u|^{2}u+V\left( x,y\right) u.
\label{eq:NLS}
\end{equation}
Here $u(x,y,t)$ is a complex-valued field, and, as indicated above, the
dispersions along the $x$- and $y$-directions have opposite signs. This
setting corresponds, for the fixed sign of the cubic nonlinearity, to
self-focusing and defocusing in the former and latter directions,
respectively. In terms of the realization in a planar optical waveguide
in the particular context of nonlinear optics, the
evolution variable $t$ is actually the propagation distance, while $x$ is
the transverse coordinate, and $y$ is the reduced temporal 
coordinate~\cite{Agrawal}. 
In this case, terms $u_{xx}$ and $u_{yy}$, with the signs adopted
in Eq.~(\ref{eq:NLS}), represent, respectively, the paraxial diffraction and
normal GVD, while the cubic term stands for the usual Kerr nonlinearity. In
what follows below, we consider the settings without the external potential,
$V(x,y)=0$, and with a 1D potential, $V(x)$, that will help to
\textquotedblleft guide\textquotedblright , and potentially stabilize the
soliton stripes.

In the absence of the potential, Eq.~(\ref{eq:NLS}) admits a solution
in the form of quasi-1D
bright-soliton stripes,
\begin{equation}
u\left( x,y,t\right) =A\,\sech\left[ A\left( x-x_{0}-vt\right) \right]
e^{i\left( vx+\mu t+\theta _{0}\right) },  \label{eq:BS}
\end{equation}
with
\begin{equation}
\mu =\frac{A^{2}-v^{2}}{2}. 
\label{mu}
\end{equation}
This is a straightforward 2D homogeneous extension of the commonly known 1D
bright soliton of the focusing NLS equation. It contains four free
parameters which determine its height (and inverse width) $A$, velocity $v$, 
initial position $x_{0}$, and initial phase $\theta _{0}$.

The study of the stability of bright soliton stripes has a time-honored
history, including some prevalent misconceptions about the type and origin
of some of its ensuing instabilities (see for instance 
Ref.~\cite{JiankeBook:2010} 
for a comprehensive review of the subject). As bright
soliton solutions are stable in 1D, it is natural to study the stability
once they are uniformly extended in 2D along the $y$ direction. In the
elliptic dispersion case [with a minus sign in front of $u_{yy}$ in 
Eq.~(\ref{eq:NLS}), 
which corresponds to the anomalous GVD in optics~\cite{Agrawal}],
the NLS is prone to collapse if the (squared) $L^{2}$ norm (alias total energy, in
terms of optics) of a localized mode exceeds the critical value, which
coincides with the corresponding
norm of the Townes solitons~\cite{sulem,Fibich}. The norm of
soliton stripes, as they extend to $y=\pm \infty $ is infinite,
hence the stripes are
always prone to the modulational instability, which is related to the
collapse-onset trend~and is called \emph{necking} 
instability~\cite{Kuznetsov:1986}. 
It eventually leads the soliton stripe to breakup into
individual lumps that are subsequently led to collapse
(individually, or after their merger with other such lumps)~\cite{review}.

On the other hand, in the framework of Eq.~(\ref{eq:NLS}), which includes
the hyperbolic dispersion, the $y$-direction is effectively (in tandem with
the nonlinearity) a \emph{defocusing} one, and, thus, it should provide a
stabilizing effect against necking. To some degree, the necking is indeed
reduced due to the defocusing nature of the sign combination of the 
$y$-dispersion (normal GVD in optics) and nonlinearity. However, as it was
detailed in Ref.~\cite{Bernard-Peli:2006} and is explained below, the bright
soliton stripe in the hyperbolic case still suffers from (relatively weak;
see below) necking instabilities. On the other hand switching from elliptic
to hyperbolic dispersion introduces strong \emph{snaking} transverse
instability on the bright-soliton 
stripe~\cite{Kuznetsov:1986,Kivshar-Peli:2000,fermions}. 
This instability was originally
studied by Zakharov and Rubenchik~\cite{Zakharov-Rubenchik:1974} ---and
later addressed in other works, see for instance 
Refs.~\cite{Peli:2001,Bernard-Peli:2006}---
using a perturbative approach for small wavenumbers
$k$ of the snaking modulation, up to the second order. The results from
these studies, naturally limited to small wavenumbers, incorrectly predicted
that the snaking instability was only present in a window starting at $k=0$
(mainly because of the introduction of a wavenumber cutoff), when it is
indeed present at all wavenumbers. These conclusions initially led to some
misconceptions like the fact that transverse instabilities were only limited (as it
occurs in the elliptic case) to a finite range of wavenumbers and, thus,
that short-wavelength transverse perturbations might be stable.
Later, numerical and analytical
studies~\cite{Cohen:1976,Saffman-Yuen:1978, Anderson:1979,Bernard-Peli:2006, Alexe}
confirmed that the snaking and necking instabilities are indeed present beyond the
window predicted in Ref.~\cite{Zakharov-Rubenchik:1974} 
(see Ref.~\cite{Peli2} for a rigorous proof of these facts). In fact, 
Ref.~\cite{Bernard-Peli:2006}, using an Evans-function 
technique~\cite{Kapitula-Sandstede:1998}, 
has clarified the origin of a secondary
instability that precludes the stabilization of short-wavelength
perturbations.

\begin{figure}[tbh]
\includegraphics[width=0.8\columnwidth]{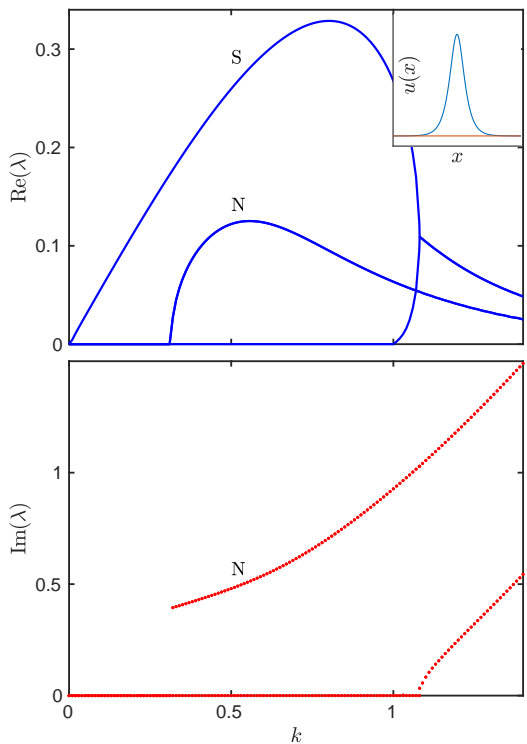}
\caption{(Color online) Real and imaginary (top and bottom panels,
respectively) of the eigenvalues associated with instabilities of the
stationary bright-soliton stripe with amplitude $A=1$ [i.e., $\protect\mu=0.5$ 
in Eq.~(\protect\ref{mu})] in the case of the hyperbolic dispersion. A
positive real part indicates instability, while the presence of an imaginary
part indicates that the instability corresponds to a quartet of eigenvalues
accounting for the oscillatory-type instability. Labels `S' and `N' denote
the snaking and necking instabilities, respectively. The inset depicts the
stationary solution $u_{0}(x)$ [see Eq.~(\protect\ref{eq:BS})] in the window 
$-10<x<+10$ and $-0.1<u<+1.2$. The corresponding perturbation eigenmodes for 
$k=0.5$ are depicted in Fig.~\protect\ref{fig:SNmodesk0.5}. }
\label{fig:spectra_sig0.0}
\end{figure}

The, previously known, instability properties for the bright-soliton stripe
in the hyperbolic-dispersion case are summarized in 
Fig.~\ref{fig:spectra_sig0.0} where the 
instability branches are denoted by `S' and
`N' for snaking and necking instabilities, respectively. The instability
spectrum can be numerically obtained by perturbing the exact stationary
solution, $u_{0}(x)$, given by Eqs.~(\ref{eq:BS}) and (\ref{mu}) with 
$x_{0}=v=\theta _{0}=0$. In what follows below, we select the stationary
solution by fixing $\mu =1/2$ in Eq.~(\ref{mu}). To perform the stability
analysis, $u_{0}$ is perturbed as~per the standard 
expression~\cite{JiankeBook:2010},
\begin{eqnarray}
u(x,y,t) &=&\left\{ u_{0}(x)+\left[ a(x)+b(x)\right] e^{\lambda t+iky}\right.
\notag \\
&&\left. +\left[ a^{\star }(x)-b^{\star }(x)\right] e^{\lambda ^{\star
}t-iky}\right\} e^{i\mu t},  \label{eq:eigen}
\end{eqnarray}
where $(\cdot )^{\star }$ stands for complex conjugation, and $\lambda $ yields
the instability growth rate (if it has a positive real part) of the
transverse modes with real wavenumber $k$. Thus, for each value of $k$, one
can compute the spectrum of eigenvalues $\lambda $, substituting expression 
(\ref{eq:eigen}) in Eq.~(\ref{eq:NLS}), linearizing it with respect to the
perturbations and solving the ensuing eigenvalue problem for $\lambda$
and the corresponding eigenvector pair $(a(x),b(x))$.

\begin{figure}[htb]
\includegraphics[width=0.55\columnwidth]{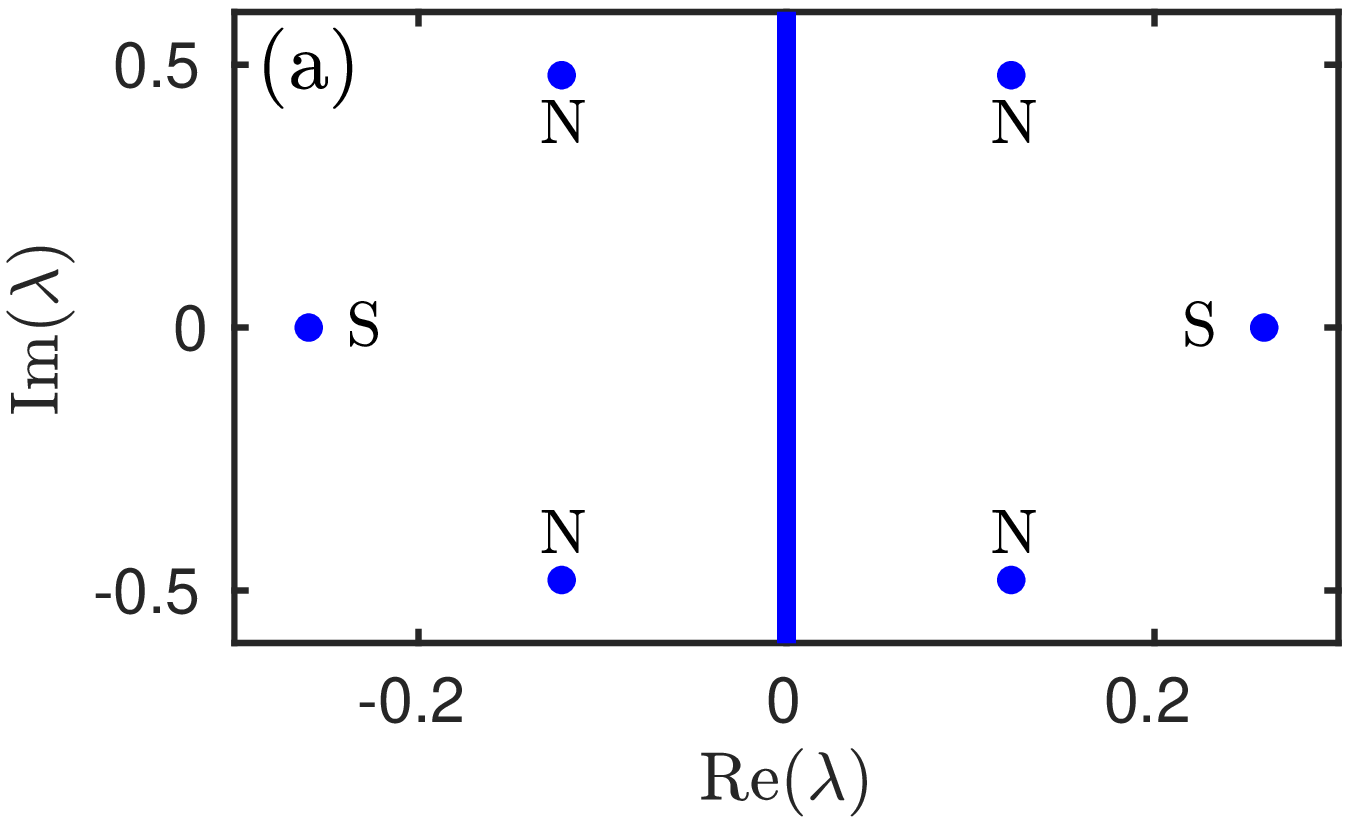} \\[2.0ex]
\includegraphics[width=0.75\columnwidth]{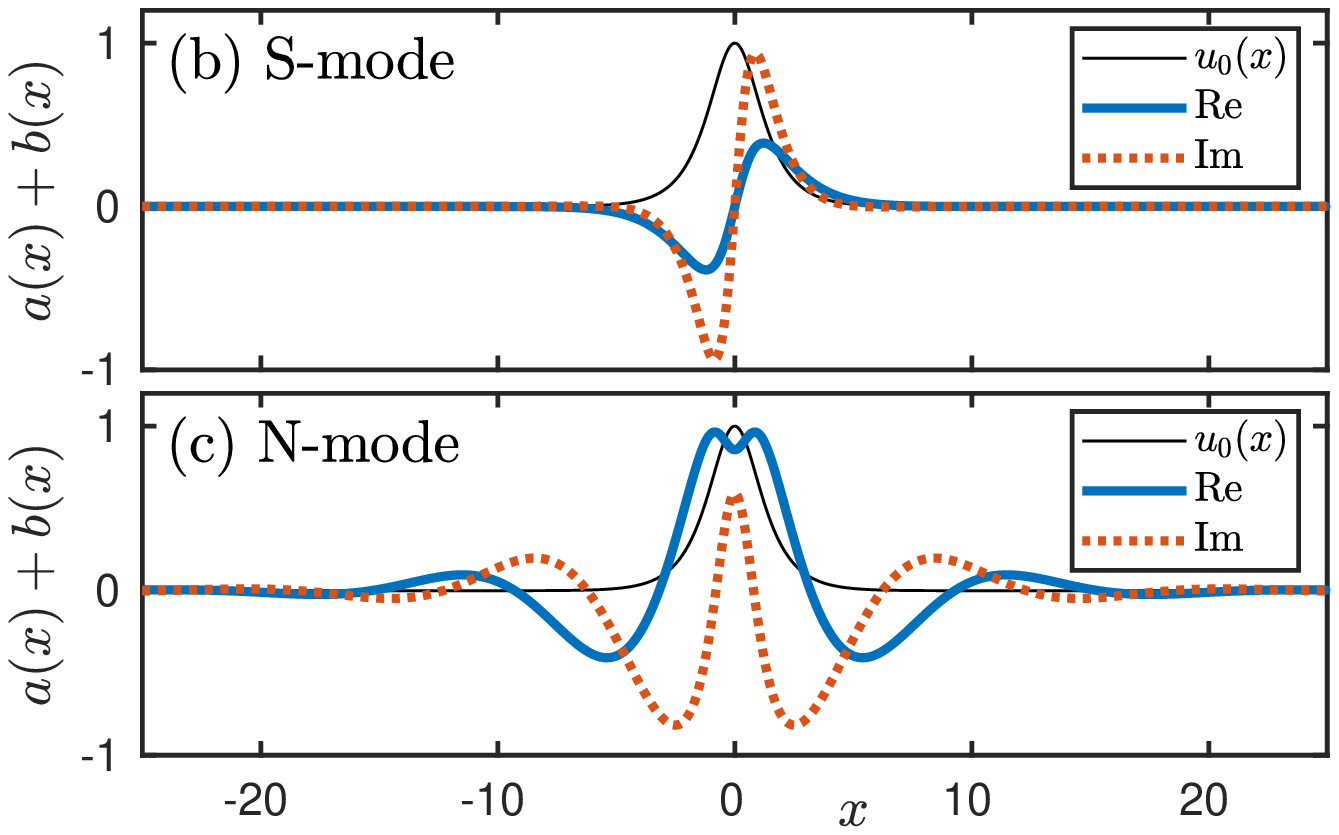} \\[2.0ex]
\includegraphics[width=1.00\columnwidth]{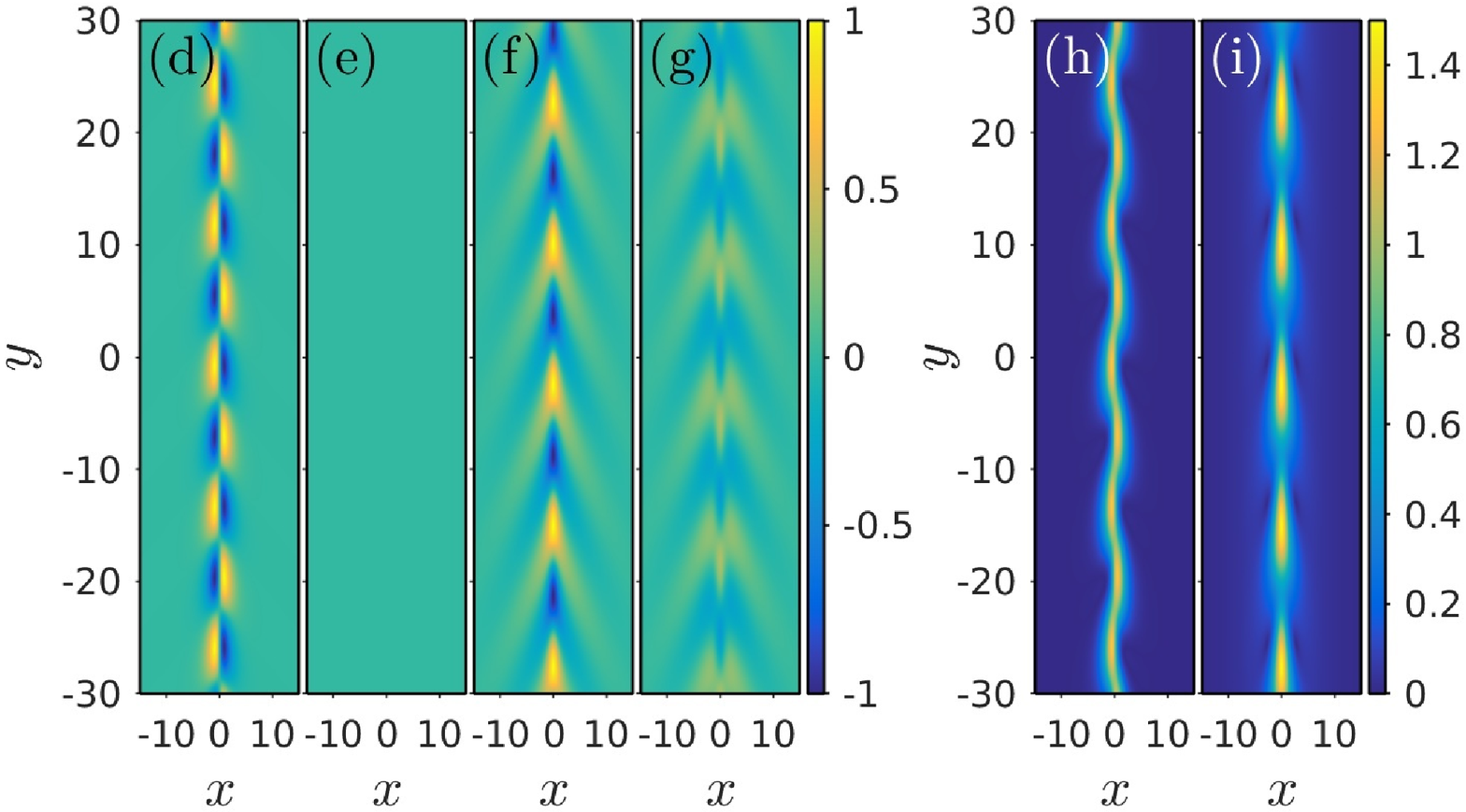}  
\caption{(Color online) 
Stability eigenvalues and eigenfunctions [the latter
correspond to the perturbation in Eq.~(\protect\ref{eq:eigen})] for $k=0.5$.
(a) The stability spectra depicting: a pair of real eigenvalues
corresponding to the snaking (S) instability, a quartet of complex
eigenvalues corresponding to necking (N) instabilities, and the continuous
spectrum consisting of purely imaginary eigenvalues. Panels (b) and (c)
depict, respectively, the $x$-dependent part of the eigenfunctions for the
snacking and necking modes (real and imaginary parts are shown by solid blue
and dashed red lines, respectively), together with the stationary steady
state $u_{0}(x)$ (the thin black line). Panels (d)--(g) depict the full
eigenfunctions in the $(x,y)$ \ plane: (d) and (e) [(f) and (g)] are the
real and imaginary parts of the snaking [necking] mode. Panels (h) and (i)
show the absolute value of the stationary state perturbed by $0.5$ times the
snaking and necking instability modes (normalized so that their maximum
absolute value is $1$), respectively. }
\label{fig:SNmodesk0.5}
\end{figure}

Further, the stability spectrum and eigenmodes corresponding to the snaking
and necking instabilities for the perturbation wavenumber $k=0.5$ are
depicted in Fig.~\ref{fig:SNmodesk0.5}. As seen in the figure, the snaking
perturbation mode, which is responsible for creating transverse undulations
of the position of the bright-soliton stripe, naturally features an \emph{odd} 
shape in the $x$-direction. On the other hand, the necking instability is,
also quite naturally, represented by an \emph{even} mode, which represents
longitudinal modulations of the local density. As mentioned above, one may
intuitively expect that the necking instability should not be present if the
model is effectively defocusing in the $y$-direction. Nonetheless, the
instability of the necking type is present, due to the fact that the
eigenvalues that give rise to this instability originate from a collision
between the eigenvalue at the edge of the continuous spectrum for small
values of $k$ and an eigenvalue bifurcating from the origin (at $k=0$), see
Refs.~\cite{Bernard-Peli:2006,Yang-Pelin} for details. The respective
necking eigenmode inherits properties of its \textquotedblleft
parent\textquotedblright\ eigenmodes leading to the bifurcation. Namely, the
mode at the edge of the continuum spectrum with small $k$ (the first
\textquotedblleft parent\textquotedblright ) is extended (periodic) along
the $x$-direction, while the mode bifurcating from the origin (the second
\textquotedblleft `parent\textquotedblright ) is an even one, which is
essentially localized in the region occupied the underlying steady state.
Therefore, the emerging necking mode is an even one, weakly localized (wider
than the stationary solution), with oscillatory tails in the $x$-direction.
These observations will be crucial when we describe the dynamics of the
bright soliton-stripe using the variational approximation in 
Sec.~\ref{sec:VA}, where this necking mode cannot be captured. 
Furthermore, the mixed
nature of the necking mode, together with the fact that the real part of its
eigenvalue is about twice as small as the one corresponding to the snaking
instability, and that necking is not present for small values of $k$, make
this mode somehow elusive when performing direct
numerical simulations (and, by the
same token, elusive in possible experiments), which might be the reason why
it was usually missed.

\section{Reduced effective equations for the soliton stripe}
\label{sec:reduced}

In this section we report two methods to derive effective reduced dynamics
of the bright-soliton stripes. The techniques, based on the
adiabatic-invariant (AI) approach~\cite{AI_us:2017,AI_us:2018} and the
variational approximation~\cite{progress}, rely on a suitable
\textquotedblleft projection\textquotedblright\ of the soliton stripe onto a
subset of collective coordinates which account for basic features of the
stripe, such as, in particular, its center position along the $x$-axis
and its dependence on the transverse variable $y$.

\subsection{The use of the adiabatic invariant (AI)}
\label{sec:AI}

We adapt here the AI reduction approach, which was developed for
dark-soliton stripes in Ref.~\cite{AI_us:2017,AI_us:2018}, to the case of
bright solitons. The method relies on using the conservation of the
Hamiltonian to derive an equation of motion for the center of
the soliton (along the $x$-direction).
In the absence of the external potential ($V=0$), the 2D
Hamiltonian corresponding to Eq.~(\ref{eq:NLS}) is
\begin{equation}
H_{\mathrm{2D}}=\frac{1}{2}\iint_{-\infty }^{\infty }\left[
|u_{x}|^{2}-|u_{y}|^{2}-|u|^{4}\right] dx\,dy.  \label{eq:H_2D}
\end{equation}
To approximate the dynamics of the soliton stripe, we consider the 1D
soliton ansatz (\ref{eq:BS}), where the position of the stripe, $X(y,t)$, is
assumed to be a function of the transverse coordinate and time:
\begin{gather}
u(x,y,t)=A\,\sech\left[ A\left( x-X(y,t)\right) \right] \times
\notag \\[0.01in]
\exp \left\{ i\left[v(x- X(y,t))+
\frac{1}{2}\left(A^{2}+v^2)\right) \,t\right] \right\} .
\label{eq:BS_AI}
\end{gather}
%
%
Within the AI approximation, we assume that deformations of the stripe
weakly affect the velocity, and do not affect
the amplitude and width of the transverse soliton's profile.
We have in mind here a nearly stationary soliton scenario, and
in assuming an invariant amplitude/width, we attempt to tackle the
phenomenon of snaking, rather than that of necking, involving
width and amplitude variation.
As we are adopting a traveling wave ansatz of the general form 
$u=f(x-X)\exp (ig(x-X))$, it immediately follows that 
$|u_{y}|=|X_{y}|\cdot |u_{x}|$, and, therefore, the Hamiltonian 
(\ref{eq:H_2D}) can be rewritten as
\begin{equation}
H_{\mathrm{2D}}=\frac{1}{2}\iint \left[
(1-X_{y}^{2}(y,t))|u_{x}|^{2}-|u|^{4}\right] dx\,dy.  \label{H}
\end{equation}
The evaluation of this expression for the stripe ansatz given by 
Eq.~(\ref{eq:BS_AI}) after identifying that $v=X_t$
and subsequent integration in the $x$-direction yields
\begin{equation*}
H_{\mathrm{2D}}=\frac{1}{2}\int_{-\infty }^{+\infty }\left[
(1-X_{y}^{2})\left( \frac{1}{3}A^{3}+AX_t^{2}\right) -\frac{2}{3}A^{3}
\right] dy.
\end{equation*}
%
%
It is important to reiterate here that the steps above rely on the AI assumption
that the velocity $v$ is only \emph{weakly}-dependent,
on $y$ (and $t$) through $v=X_t(y,t)$.

We now use the fact that the Hamiltonian is conserved, i.e., $dH/dt=0$, to
obtain
\begin{equation}
\int_{-\infty }^{+\infty }\left[ (1-X_{y}^{2})\left( \frac{1}{3}
A^{3}+AX_{t}^{2}\right) -\frac{2}{3}A^{3}\right] _{t}dy=0.  \label{integral}
\end{equation}
We apply the integration by parts in Eq.~(\ref{integral}), assuming that the
variation of the stripe vanishes at $y\rightarrow \pm \infty $,
\begin{gather*}
\!\!\!\!\int_{-\infty }^{+\infty }\left[ X_{yy}\left( \frac{A^{3}}{3}
+AX_{t}^{2}\right) \right.  \\[0.01in]
\left. \phantom{\frac{1}{3}}+2X_{y}X_{yt}X_{t}-(1-X_{y}^{2})X_{tt}\right]
\!X_{t}\,dy=0,~~~
\end{gather*}
from where it follows that the reduced AI equation for the soliton's transverse
position is
\begin{equation}
(1-X_{y}^{2})X_{tt}=X_{yy}\left( \frac{A^{3}}{3}+AX_{t}^{2}\right)
+2X_{y}X_{yt}X_{t}.  \label{eq:AI}
\end{equation}

If we consider small perturbations from the initial straight soliton stripe,
the linearization of Eq.~(\ref{eq:AI}) yields
\begin{equation*}
X_{tt}=\frac{A^{3}}{3}X_{yy},
\end{equation*}
which recovers the well-known snaking instability of the bright soliton
stripe~\cite{Kuznetsov:1986}, which is described by the following dispersion
relation, assuming $X\sim \exp (i(ky-\omega t))$,
\begin{equation}
\omega ^{2}+\frac{A^{2}}{3}k^{2}=0.  \label{eq:AI_disp_rel}
\end{equation}
We stress that the main result of the AI approach is not to predict the
linear instability of the stripe. The AI goes further, as Eq.~(\ref{eq:AI})
should in principle allow one to follow the \emph{nonlinear} evolution of
the snaking past its initial instability. In Sec.~\ref{sec:numerics} below
we will compare the AI prediction, based on Eq.~(\ref{eq:AI}), with the full
numerically simulated evolution of the original NLS equation (\ref{eq:NLS}).

\subsection{The variational approximation (VA)}
\label{sec:VA}

\subsubsection{The full VA}
\label{sec:fullVA}

As presented above, the AI reduction approach, has an intrinsic shortcoming:
as we only use a \emph{single} conserved quantity (the Hamiltonian),
this approach can only produce a \emph{single} equation of motion. 
In the case of the dark-soliton stripe, as put forward in 
Refs.~\cite {AI_us:2017,AI_us:2018}, the single equation, 
governing the transverse shift of the dark-soliton's center, 
is sufficient to describe its dynamics ---as the width of the
dark soliton is determined by the height of the constant background into
which the dark soliton is embedded and the soliton speed.
However, in the present case, the bright soliton, in principle,
has more degrees of freedom (through the potential variation of
its amplitude) and, thus, it will be relevant to
develop an approach that can predict the evolution of these degrees of
freedom along the stripe. 
In principle, a more general AI description could be obtained by using
the Hamiltonian formulation together with its corresponding
\emph{canonical variables}.
However, at this point, it remains elusive in our setting precisely 
what is the right choice of canonical variables that could describe the 
soliton filament's dynamics, so as to use Hamilton's equations around
such a canonical formulation. 
Therefore, instead of using a Hamiltonian formulation, we turn to a more
straightforward Lagrangian one, enabling the derivation of Euler-Lagrange
equations for both the center position and the width (or equivalently
amplitude) of the solitary wave.
In that vein, we adapt a quasi-2D VA methodology, as
proposed in Ref.~\cite{Malomed92}, to develop the full description of the
dynamics of the bright-soliton stripe. The VA is based on the Lagrangian of
Eq.~(\ref{eq:NLS}) (this time, it includes the external potential $V$):

\begin{eqnarray}
L &=&\iint_{-\infty }^{\infty }\left[ \frac{i}{2}\left( u^{\ast
}u_{t}-uu_{t}^{\ast }\right) -\frac{1}{2}|u_{x}|^{2}+\frac{1}{2}
|u_{y}|^{2}\right.   \notag \\[0.01in]
&&\left. +\frac{1}{2}|u|^{4}-V(x,y)|u|^{2}\right] dx\,dy.  \label{eq:Lag}
\end{eqnarray}
Based on the exact bright-soliton solution of Eq.~(\ref{eq:BS}), we
introduce the following 2D ansatz for the respective stripe:
\begin{eqnarray}
&&u\left( x,y,t\right) =A\left( y,t\right) \sech\left[ A\left( y,t\right)
\left( x-\xi \left( y,t\right) \right) \right] 
\times   
\notag \\[0.01in]
&&
\exp\left\{i\left[ v(y,t)(x-\xi (y,t))+\theta (y,t)\right] \right\},
\label{eq:VA_ansatz}
\end{eqnarray}
where, in contrast with the AI ansatz~(\ref{eq:BS_AI}), we now have four $t$- and 
$y$-dependent parameters. These four variables represent the
amplitude (or inverse
width) $A(y,t)$ of the soliton, its location $\xi (y,t)$ [which plays the
same role as $X(y,t)$ in the AI ansatz (\ref{eq:BS_AI})], the $x$-velocity 
$v(y,t)$, and the phase $\theta (y,t)$ [the solution with 
$A\left( y,t\right)=$constant, $v(y,t)=$constant, 
$\xi (y,t)=vt$ and $\theta (y,t)=\frac{1}{2} v^{2}t+\frac{1}{2}A^{2}t+\theta _{0}$ 
amounts to the exact bright-soliton
stripe in the absence of any external potential].

Substituting the VA ansatz (\ref{eq:VA_ansatz}) in the
Lagrangian (\ref{eq:Lag}), we
perform its $x$-direction integration
for the trial function (\ref{eq:BS}), which yields
\begin{equation}
L=\int_{-\infty }^{+\infty }{\mathcal{L}}\,dy,  \label{L}
\end{equation}
with the Lagrangian density
\begin{eqnarray}
\mathcal{L} &\!\!=\!\!&-2A\left( \theta _{t}-v\xi _{t}\right) +\frac{A^{3}}{3}
(\xi _{y}^{2}+1)-Av^{2}+\frac{12+\pi ^{2}}{36}\frac{A_{y}^{2}}{A}  
\notag \\[0.01in]
&&+\frac{\pi ^{2}}{12}\frac{v_{y}^{2}}{A}+A\left( v\xi _{y}-\theta
_{y}\right) ^{2}-V^{\mathrm{eff}}(A,\xi ),  \label{LL}
\end{eqnarray}
where we do not explicitly write the dependence on $(y,t)$ of the dynamical
variables, and an effective (averaged over the ansatz) potential is
introduced:
\begin{equation}
V^{\mathrm{eff}}(A,\xi )=\int_{-\infty }^{+\infty }V(x,y)\,h(A,x-\xi )\,dx,
\label{eq:Veff}
\end{equation}
where $h(A,r)\equiv A^{2}\sech^{2}\left( Ar\right) $. Note that the
effective potential depends on time and the longitudinal coordinate $y$ through
its dependence on $A(y,t)$ and $\xi (y,t)$.

Now, following the standard VA method, we perform variations of parameters 
$\theta $, $v$, $\xi $, and $A$ to produce the following Euler-Lagrange
equations. First, it is
\begin{equation}
\delta \theta :A_{t}-\left[ A\left( \theta _{y}-v\xi _{y}\right) \right]
_{y}=0,  \label{eq9a}
\end{equation}
which is tantamount to the continuity equation, which secures the
conservation of the squared $L^2$ norm (alias \textit{total energy}, in terms of
the optical realization, which is different from the above-mentioned
Hamiltonian),
\begin{equation}
E=\iint |u(x,y)|^{2}dx\,dy,  \label{E}
\end{equation}
which diverges (linearly, as a function of
the domain size)  for the stripes. From Eq.~(\ref{eq:BS}) 
it follows $\int_{-\infty }^{+\infty }|u(x,y,t)|^{2}dx=2A$, hence
the integration of Eq.~(\ref{eq9a}) with respect to $y$ produces:
\begin{equation}
{\partial _{t}}\iint |u(x,y,t)|^{2}dx\,dy=2A\left( \theta _{y}-v\xi
_{y}\right) |_{y=-\infty }^{y=+\infty },
\notag
\end{equation}
which corresponds to the energy flux along the $y$-axis.

Next, the variation of the Lagrangian produced by Eqs.~(\ref{L}) and 
(\ref{LL}) with respect to $v$ and $\xi $ yields
\begin{equation}
\delta v:\xi _{t}=v-\xi _{y}\left( v\xi _{y}-\theta _{y}\right) +\frac{\pi
^{2}}{12A}\left( \frac{v_{y}}{A}\right) _{y},  \label{eq9b}
\end{equation}
\begin{equation}
\delta \xi :v_{t}=\frac{-1}{3A}\left( A^{3}\xi _{y}\right) _{y}-v_{y}\left(
v\xi _{y}-\theta _{y}\right) -\frac{1}{2A}V_{\xi }^{\mathrm{eff}}(A,\xi ),~~
\label{eq10}
\end{equation}
where Eq.~(\ref{eq9a}) was used to simplify Eq.~(\ref{eq10}), and the
subscript in $V_{\xi }^{\mathrm{eff}}\equiv \partial _{\xi }V^{\mathrm{eff}}$
denotes the derivative with respect to $\xi $ once the explicit dependence
on $A(y,t)$ and $\xi (y,t)$ in $V^{\mathrm{eff}}$ has been introduced.
Finally, the variation of the Lagrangian with respect to $A$ produces the
following equation,
\begin{gather}
-\frac{12+\pi ^{2}}{36}\left( \frac{A_{yy}}{A}-\frac{A_{y}^{2}}{2A^{2}}
\right) =\theta _{t}-v\xi _{t}-\frac{A^{2}}{2}\left( \xi _{y}^{2}+1\right)+\frac{v^2}{2}
\notag \\[0.01in]
+\frac{\pi ^{2}}{24}\frac{v_{y}^{2}}{A^{2}}-\frac{1}{2}\left( v\xi
_{y}-\theta _{y}\right) ^{2}+\frac{1}{2}V_{A}^{\mathrm{eff}}(A,\xi ),
\label{A}
\end{gather}
where $V_{A}^{\mathrm{eff}}\equiv \partial _{A}V^{\mathrm{eff}}$. In 
combination with Eq.~(\ref{eq9b}), Eq.~(\ref{A}) yields
\begin{gather}
\displaystyle\displaystyle\displaystyle\theta _{t}=\frac{1}{2}v^{2}-\frac{1}{2}
\left( v^{2}\xi _{y}^{2}-\theta _{y}^{2}\right) +\frac{\pi ^{2}v}{12A}
\left( \frac{v_{y}}{A}\right) _{y}-\frac{\pi ^{2}}{24}\frac{v_{y}^{2}}{A^{2}}
~~~~~~  \notag \\[0.01in]
\displaystyle\displaystyle\displaystyle-\frac{12+\pi ^{2}}{36}\left( 
\frac{A_{yy}}{A}-\frac{A_{y}^{2}}{2A^{2}}\right) +\frac{A^{2}}{2}\left( \xi
_{y}^{2}+1\right) -\frac{1}{2}V_{A}^{\mathrm{eff}}(A,\xi ).  \label{eq12}
\end{gather}

Thus Eqs.~(\ref{eq9a}), (\ref{eq9b}), (\ref{eq10}), and (\ref{eq12}) provide
a full approximation for the spatio-temporal evolution of the bright-soliton
stripe's parameters $A$, $\xi $, $v$ and $\theta $, respectively. 
In Sec.~\ref{sec:numerics} 
we compare this approximation to the respective numerical
solution of Eq.~(\ref{eq:NLS}).

\subsubsection{Slowly varying solutions}
\label{sec:VA_slowly}

To cast the VA in a more explicit form, we here adopt additional
assumptions, neglecting certain terms in Eqs.~(\ref{eq9a}), (\ref{eq9b}), 
(\ref{eq10}) and (\ref{eq12}). Namely, we focus on the slow evolution of the
soliton-stripe's parameters, treating the first derivatives $v_{y}$, $A_{y}$,
etc.~as first-order small quantities, and neglecting higher-order terms,
such as $v_{y}^{2}$, $v_{y}\,w_{y}$, etc.,~with the notable exception of
those corresponding to position $\xi (y,t)$ which are associated with the snaking
dynamics. In the framework of this slowly-varying assumption, the set of
equations (\ref{eq9a}), (\ref{eq9b}), (\ref{eq10}) and (\ref{eq12}) reduce
to the following nonlinear system of simplified coupled equations for the
stripe's parameters:
\begin{eqnarray}
A_{t}&=&  A\left( \theta _{yy}-v\xi _{yy}\right) ,  \label{eq:A_slow} \\
\xi _{t}&=&  v+\frac{\pi ^{2}}{12A^{2}}v_{yy},  \label{eq:xi_slow} \\
v_{t}&=&  -\frac{A^{2}}{3}\xi _{yy}-\frac{1}{2A}V_{\xi }^{\mathrm{eff}},
\label{eq:v_slow} \\
\theta _{t}&=&  \frac{v^{2}}{2}\!+\!\frac{\pi ^{2}v}{12A^{2}}v_{yy}\!-\!
\frac{12+\pi ^{2}}{36}\frac{A_{yy}}{A}\!+\!\frac{A^{2}}{2}\!-\!\frac{1}{2}
V_{A}^{\mathrm{eff}}.~~~~~~  \label{eq:theta_slow}
\end{eqnarray}
In particular, Eq.~(\ref{eq:xi_slow}) provides the first order correction,
$\sim $ $v_{yy}$, to the lowest-order relation between the position and
velocity, $\xi _{t}=v$. Actually, this correction is the improvement
provided by the VA in comparison to the (simplest variant
developed above of the) AI-based approach.

Note also that, according to the slowly-varying assumption, 
Eqs.~(\ref{eq:xi_slow}) and (\ref{eq:v_slow}) for the position and velocity decouple
from the spatial derivatives of the other two variables, the stripe's
amplitude and phase. This observation implies that, for the slow transverse
stripe's dynamics, the snaking (associated with $\xi $ and $v$) is only
weakly coupled to necking (associated with $A$ and $\theta $). 
In Sec.~\ref{sec:numerics}, we corroborate this conclusion by means 
of the comparison to numerically generated results.

\begin{figure}[tbh]
\includegraphics[width=0.85\columnwidth]{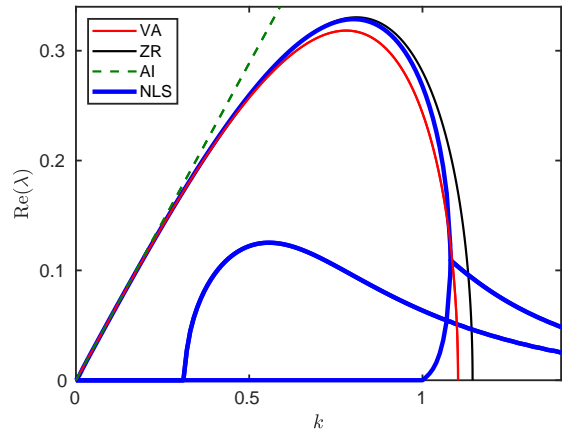}
\caption{(Color online) The real part of eigenvalues $\protect\lambda $ for
small perturbations, added to the bright-soliton stripe, as a function of
wavenumber $k$. Thick blue curves represent the full numerical results (the
same as displayed in Fig.~\protect\ref{fig:spectra_sig0.0}). 
The thin red (VA),
dashed green (AI),
and thin black (ZR) 
curves depict dispersion relations 
(\protect\ref{eq:VA_disp}), 
(\protect\ref{eq:AI_disp_rel}), and
(\protect\ref{eq:ZR})
predicted by the VA, AI, and perturbation approaches respectively. 
}
\label{fig:spectra_AI_VA}
\end{figure}

\subsubsection{The linear-stability analysis}
\label{sec:VA_stab}

In the presence of a localized external potential in the $x$ direction 
[$V(x,y)=V(x)$] in Eq.~(\ref{eq:NLS}), with a minimum at $x=0$, we
can readily expand the potential as
$V^{\mathrm{eff}}=V^{\mathrm{eff}}\left( A,0\right) +\frac{1}{2}
V_{\xi \xi }^{\mathrm{eff}}\left( A,0\right) \xi ^{2}$ for small values of 
$\xi $, as the first derivative of the effective potential vanishes, $V_{\xi
}^{\mathrm{eff}}\left( A,0\right) =-\int_{-\infty }^{+\infty }V\left(
x\right) h_{x}\left( A,x\right) dx=0$, due to the parities of the profiles 
$(x)$ and $h(A,x)$. Therefore, for a fixed value of $A$, the substitution of
planar waves in Eqs.~(\ref{eq:xi_slow}) and (\ref{eq:v_slow}), $\xi
(y,t)=Be^{i\left( ky-\omega t\right) }$ and $v(y,t)=Ce^{i\left( ky-\omega
t\right) }$ yields the dispersion relation,
\begin{equation}
\omega ^{2}+\left( \frac{k^{2}A^{2}}{3}-\frac{V_{\xi \xi }^{\mathrm{eff}}
\left( A,0\right) }{2A}\right) \left( 1-\frac{k^{2}\pi ^{2}}{12A^{2}}
\right) =0  \label{eq:VA_disp}
\end{equation}
[the linearization of full VA equations~(\ref{eq9b}) and (\ref{eq10}) leads
to the same result]. 
%
%
For instance, in the case of a delta-functional external potential, with 
$V(x)=-\epsilon \delta (x)$, $V_{\xi \xi }^{\mathrm{eff}}\left( A,0\right)
=2\epsilon A^{4}$, the VA predicts stabilization of the snaking
perturbations at $k^{2}>3\epsilon A$.
It is important to note that the dispersion relation in Eq.~(\ref{eq:VA_disp})
gives a global approximation for the instability growth rates as we did
not assume any condition on the smallness of the wavenumber $k$. 
This is in contrast
with previous results found by Zakharov and Rubenchik (ZR) using a fourth 
order perturbation analysis in the wavenumber, which, in the absence of 
external potential and for unit amplitude, i.e.~$A=1$, 
yields~\cite{Zakharov-Rubenchik:1974, Bernard-Peli:2006, Kivshar-Peli:2000}
\begin{equation}
\label{eq:ZR}
\omega^2 + \frac{k^2}{3} \left[1-\frac{1}{3}\left(\frac{\pi^2}{3}-1\right)k^2\right] = 0.
\end{equation}
%

\begin{figure}[tbh]
\includegraphics[width=1.0\columnwidth]{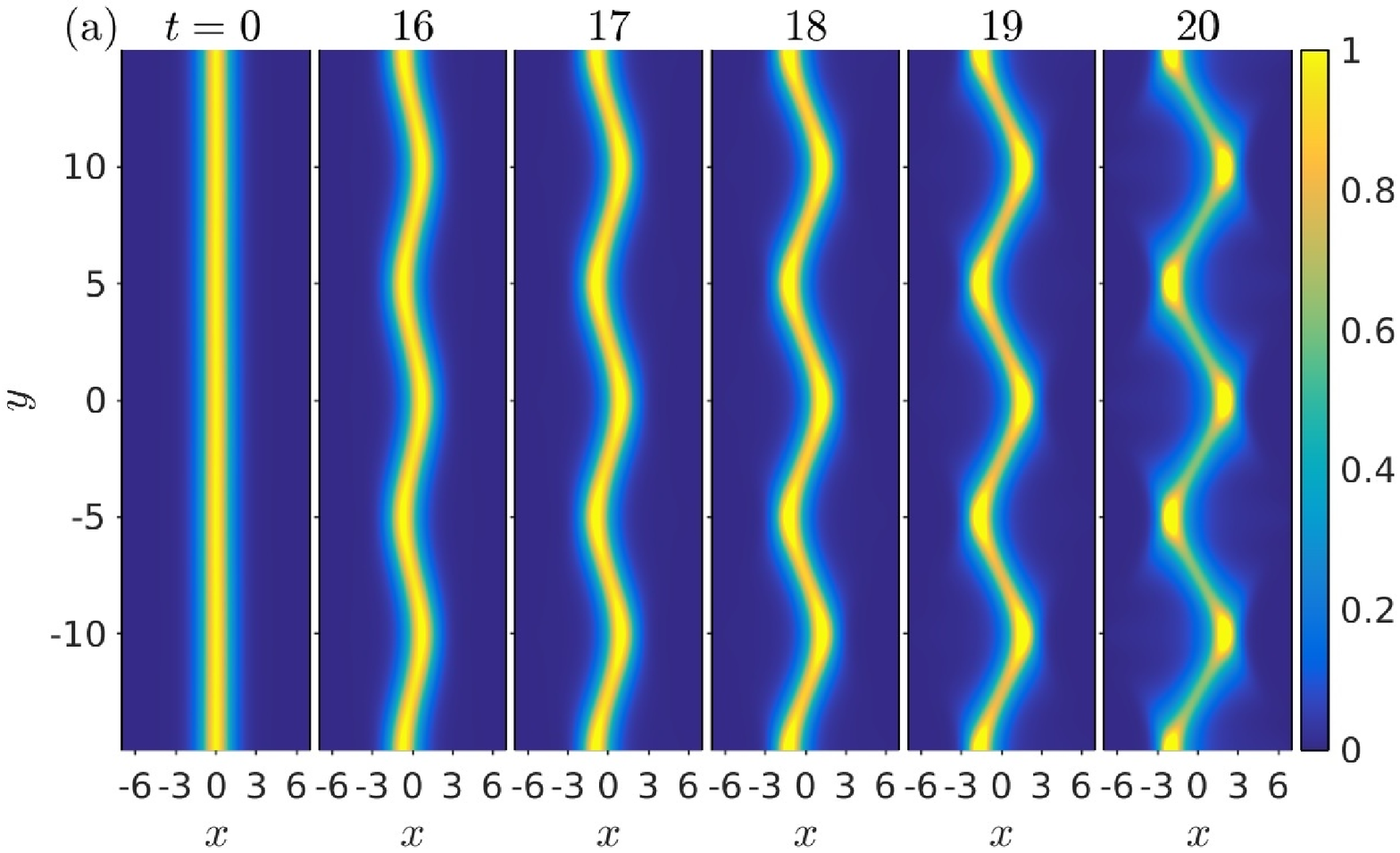}
\\[2.0ex]
\includegraphics[width=1.0\columnwidth]{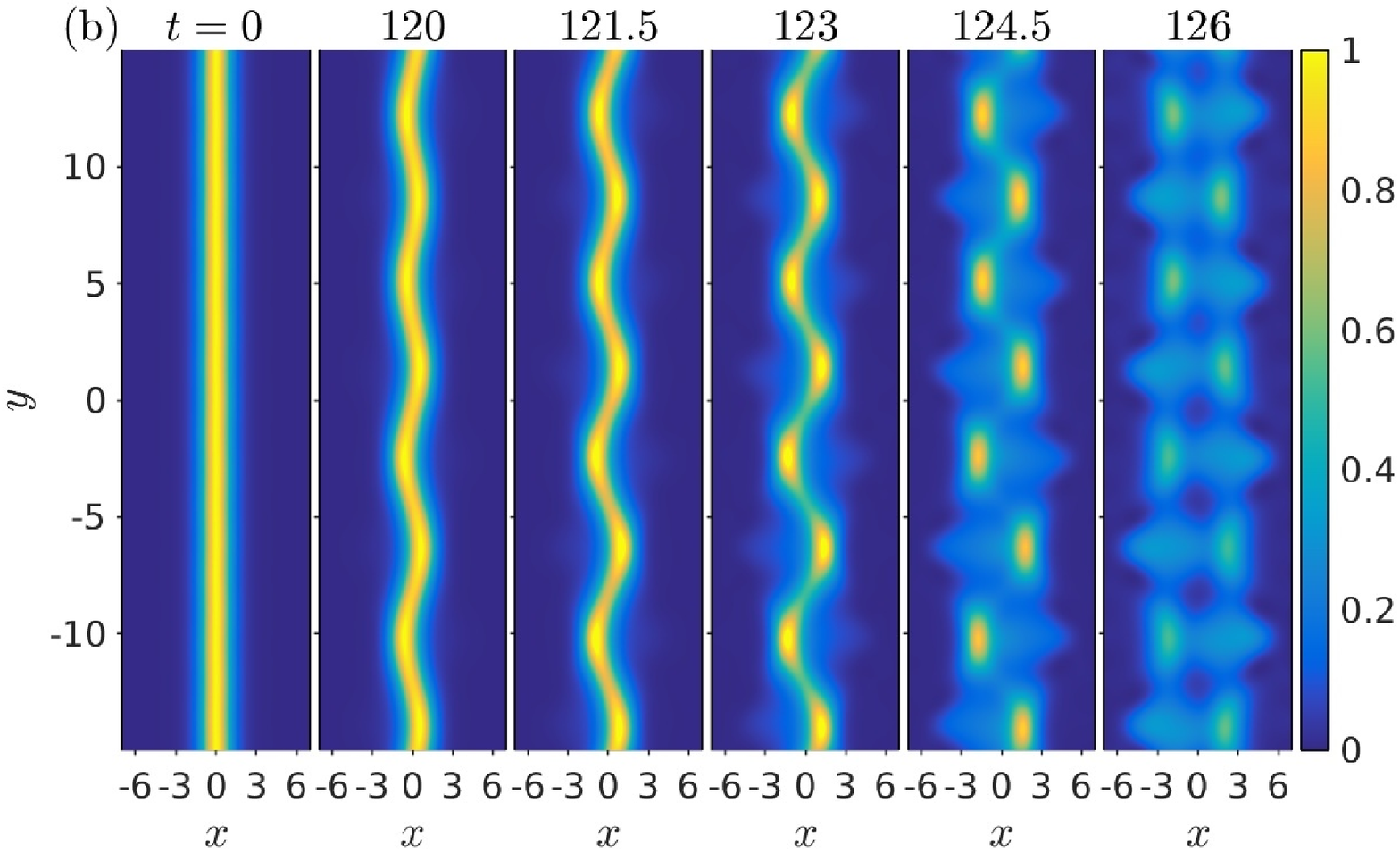}
\\[1.0ex]
\includegraphics[width=0.490\columnwidth]{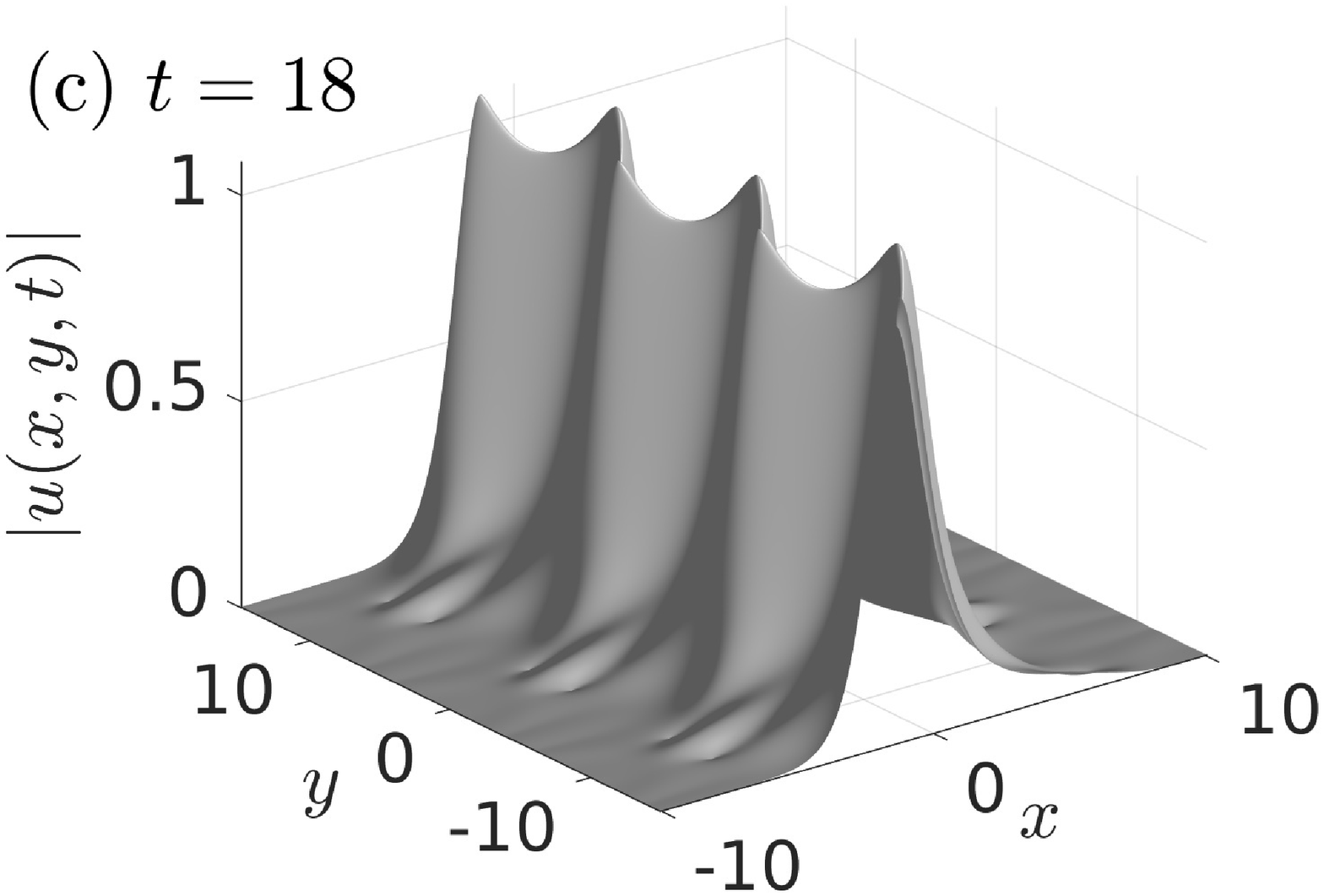}
\includegraphics[width=0.490\columnwidth]{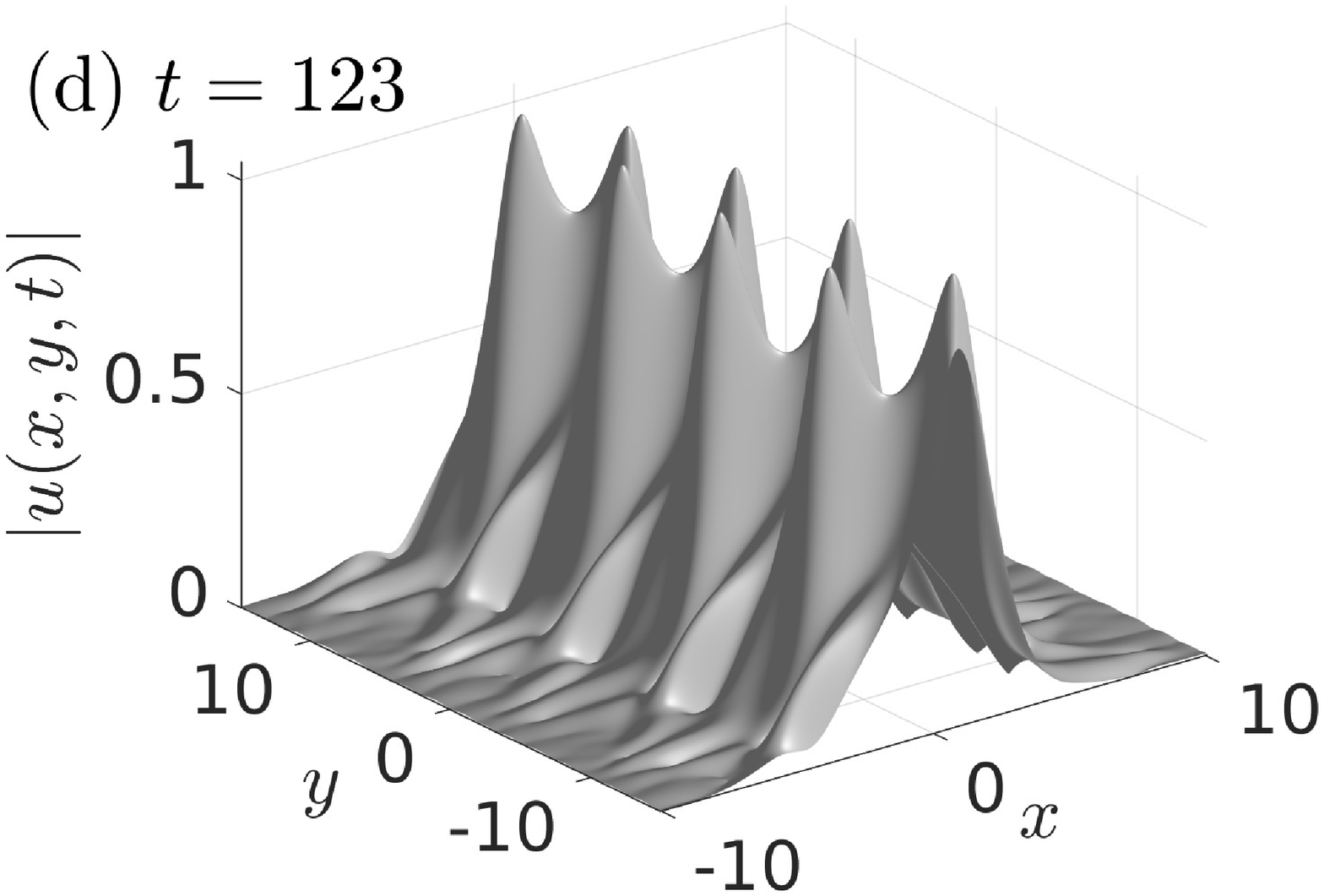}
\caption{(Color online) 
Dynamics induced by snaking instability for (a) long and
(b) short wavelengths.
The perturbation was initialized by a snaking with amplitude 0.01
and wavenumber 
(a) $k=k_1=\pi/5\simeq0.6283$ ($t=18$) and
(b) $k=k_2=2\pi/3\simeq2.0944$ ($t=123$).
Panels (c) and (d) depict a snapshot of $|u(x,y,t)|$ for $k=k_1$
and $k=k_2$, respectively. Notice the deviation of the
latter from a sech-like profile.
}
\label{fig:k_gt_kcr}
\end{figure}

Figure~\ref{fig:spectra_AI_VA} shows the comparison of the
linear-instability spectra, as predicted by the AI and VA approaches,
namely Eqs.~(\ref{eq:AI_disp_rel}) and (\ref{eq:VA_disp}),
together with the perturbative result of Eq.~(\ref{eq:ZR}),
against the full numerical solution of Eq.~(\ref{eq:NLS})
in the absence of the external potential.
As expected, all these spectra approximations asymptotically 
coincide in the $k\rightarrow 0$ limit.
It is relevant to mention that Ref.~\cite{Kivshar-Peli:2000} also 
considered this $k\rightarrow 0$ limit where effective equations for the
dynamics of the bright stripe were obtained (see Eq.~(5.36) in 
Ref.~\cite{Kivshar-Peli:2000}).
However, the VA and ZR linear spectra remain approximately valid at larger 
$k$ as well, contrary to what is the case for the simplified
AI model, which is only valid for small $k$.
Also, conceived as a global method (i.e., independent of $k$), the VA
seemingly gives a better approximation than the ZR perturbative
approach of Eq.~(\ref{eq:ZR}) for wavenumbers close to the
critical threshold $k=k_{\rm cr}\simeq 1.08$.
However, the VA fails to capture the secondary instability of the snaking
branch which the numerical solutions reveal past $k=1$. This deficiency of
the VA is explained by the fact that the secondary bifurcation of the
snaking branch at $k>1$ involves (as it is also true for the bifurcation of
the necking branch) a mode belonging to the continuum spectrum, with slower
decaying oscillatory tails, that the VA is not designed to capture in
the context of the ansatz of Eq.~(\ref{eq:VA_ansatz}).
In Fig.~\ref{fig:k_gt_kcr} we present two examples corresponding to the
destabilization dynamics for the bright soliton stripe due to snaking
for long and short wavelengths.
As it can be noticed from panels (a) and (c), the destabilization for long
wavelengths, corresponding to small values of $k$ before the bifurcation
at approximately $k=k_{\rm cr}\simeq 1.08$, keeps the transverse profile of the
stripe close to a unimodal (sech-like) hump.
On the other hand, see panels (b) and (d), for short wavelengths,
$k>k_{\rm cr}$; here, the transverse profile develops strong tails that
cannot be captured by the sech-shaped ansatz profile.

In Sec.~\ref{sec:potential} we report results obtained in the presence of
the external potential. Of particular interest is the eventual suppression
of different types of the instability of the soliton stripe by appropriately
designed potentials, at different wavenumbers $k$.

\section{Numerical results: soliton-stripe dynamics}
\label{sec:numerics}

We now aim to verify the validity of the results predicted by the AI and
(especially so) the VA
method, by going beyond the prediction for the stability of the
bright-soliton stripe. Indeed, as both the AI and the VA generate
approximate (reduced) equations of motion for the dynamics of the stripe,
the reduced equations may be valid not just for small perturbations around
the steady state, but also for large (nonlinear) perturbations. However,
for the reasons explained previously,
these approximate methods fail to reveal the necking
dynamics in the case of the hyperbolic dispersion, therefore below we focus
on the snaking dynamics. It is relevant to stress that, as the snaking
perturbations have a relatively large instability growth rate, in comparison
to necking, it is expected that the snaking dynamics ought to dominate over
necking. Below we numerically integrate the underlying equation (\ref{eq:NLS}) 
by means of a combination of the standard second-order finite-difference
algorithm in space and fourth-order Runge-Kutta scheme in time. The same
results have also been produced using spectral discretization in space. As
concerns equations produced by the AI and VA methods, they were solved
numerically by dint of standard spectral methods in space together with
Runge-Kutta integration in time.

\begin{figure}[tbh]
\hspace{0.06cm} 
\includegraphics[width=0.47\columnwidth]{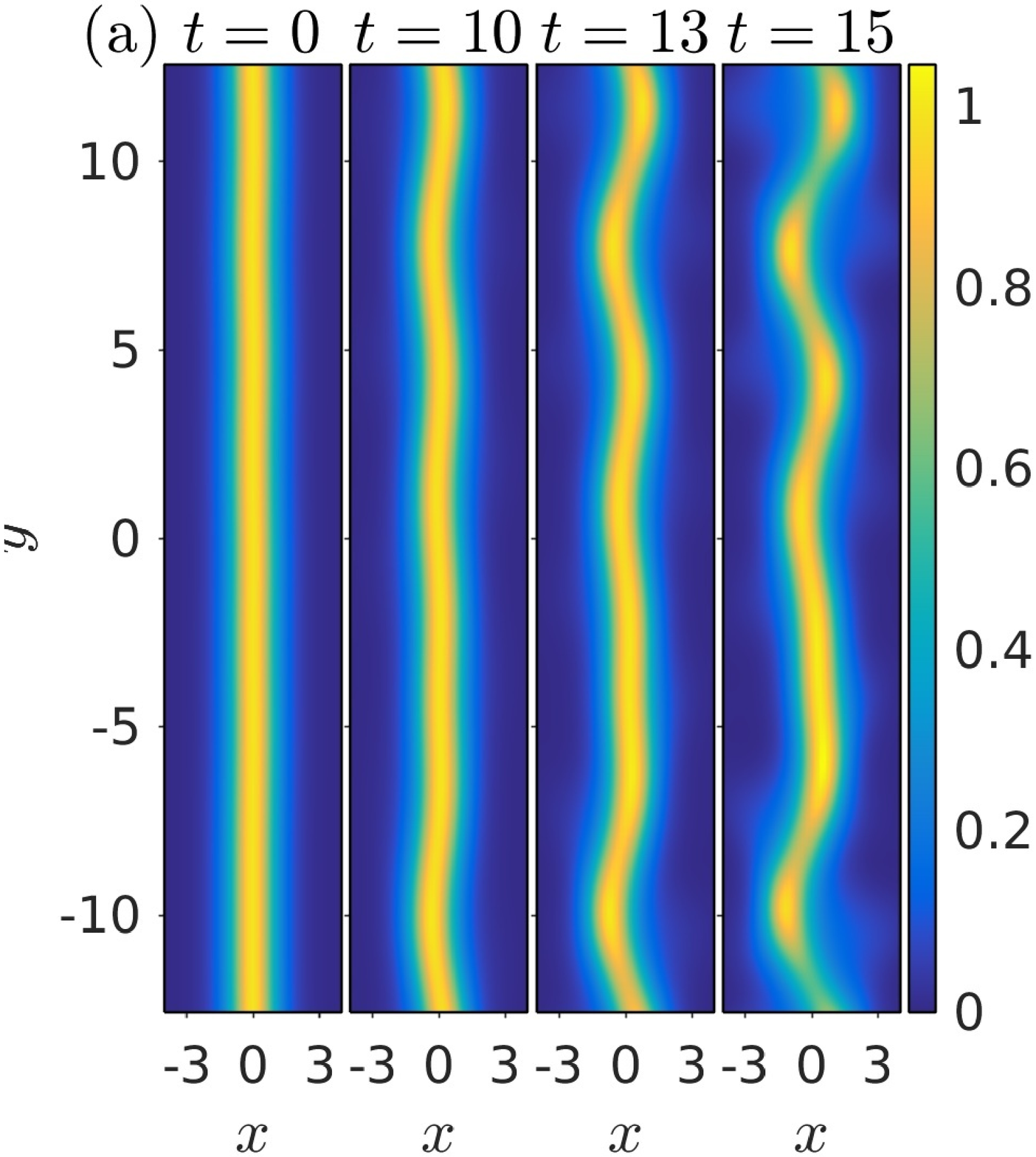} \, 
\includegraphics[width=0.47\columnwidth]{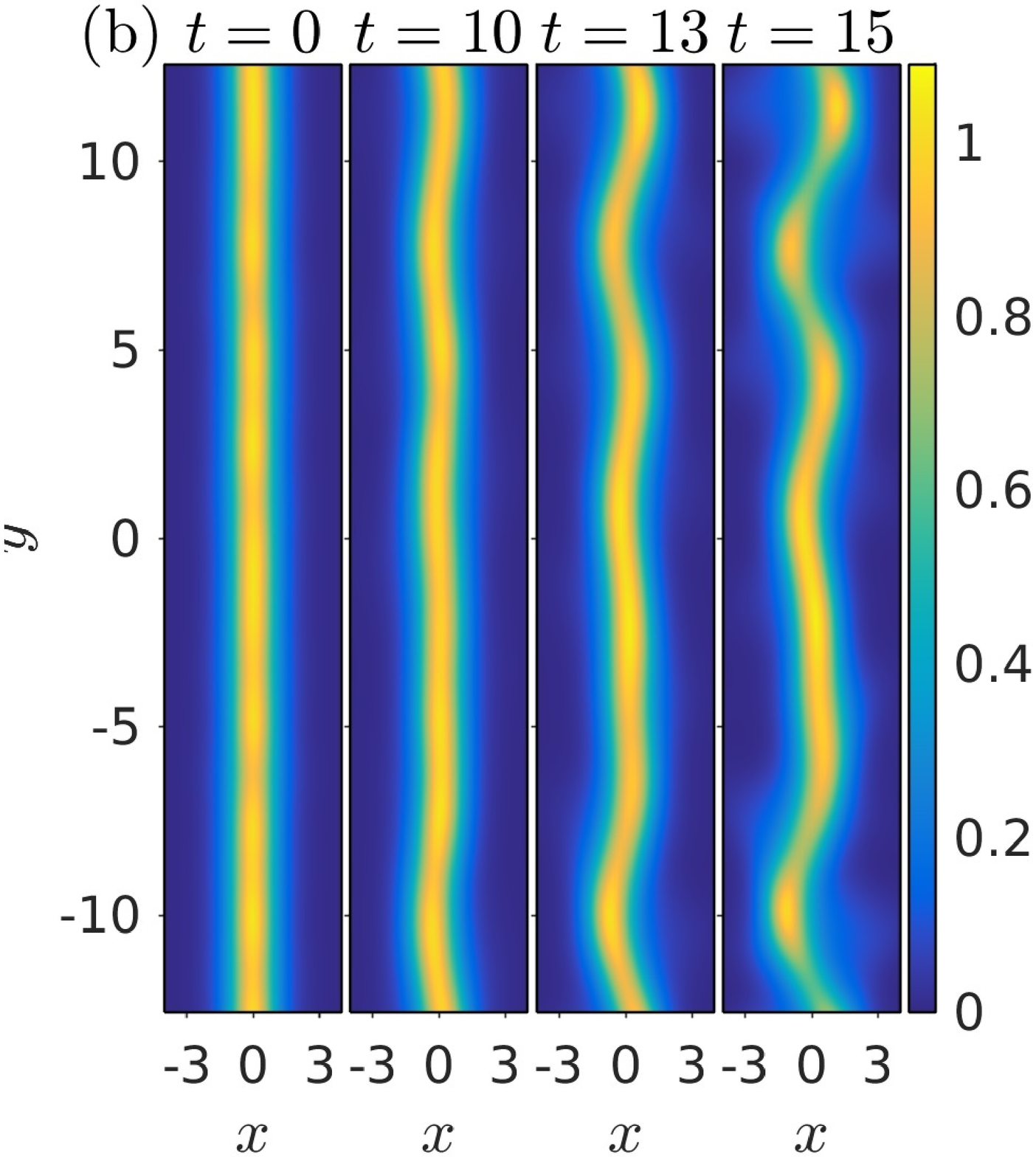} 
\\[1.0ex]
\includegraphics[width=0.49\columnwidth]{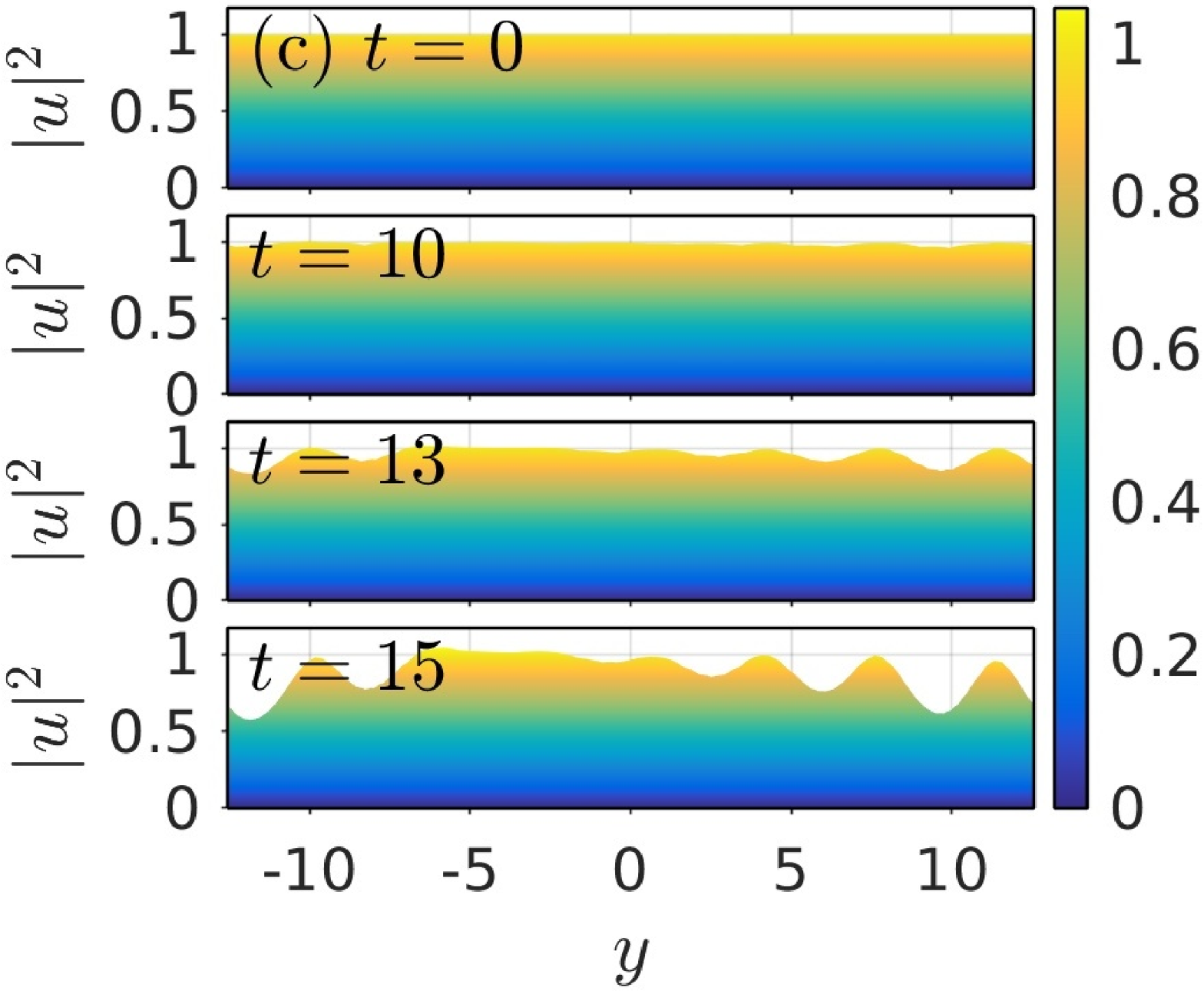} 
\includegraphics[width=0.49\columnwidth]{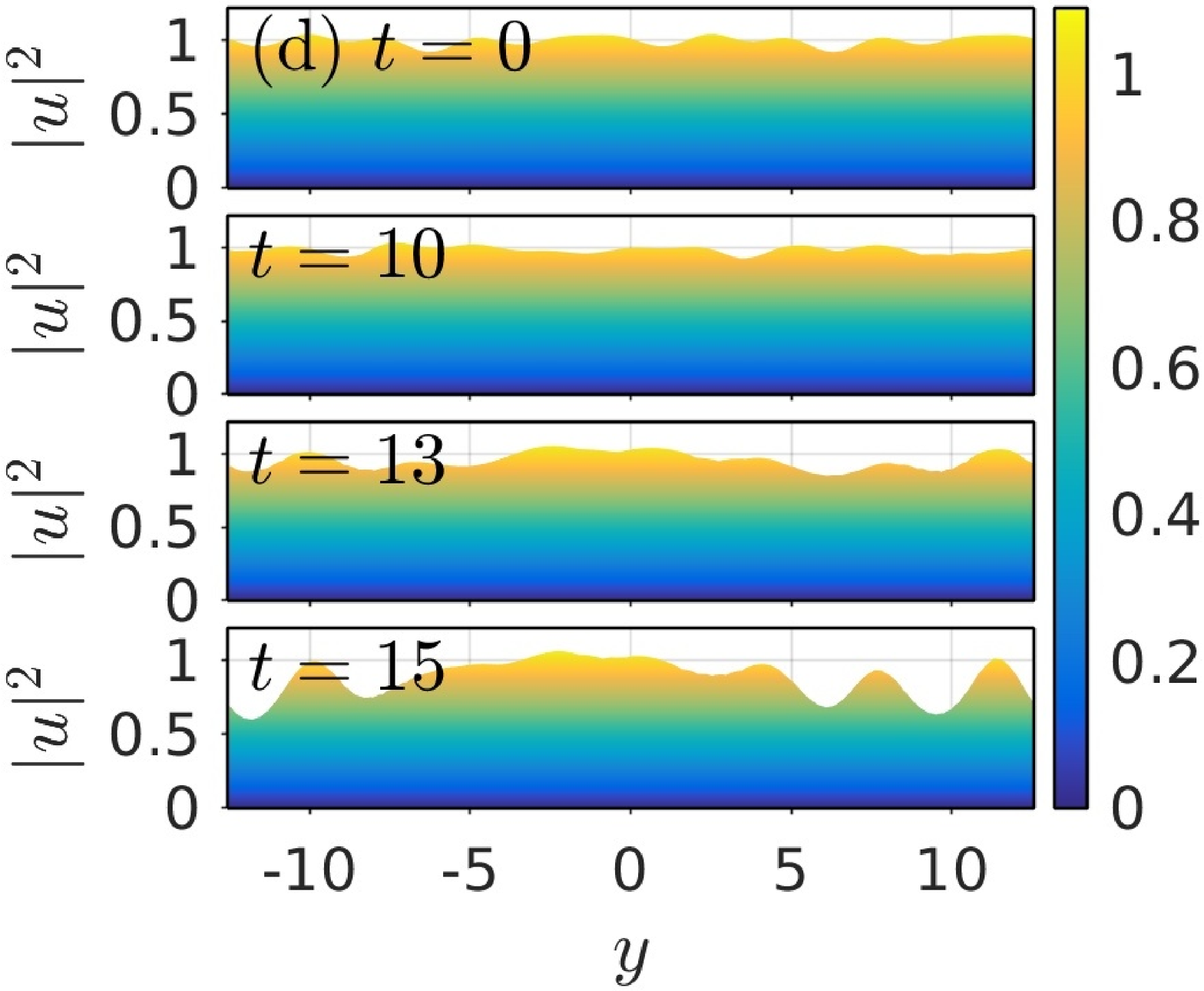}
\\[1.0ex]
\includegraphics[height=4.2cm]{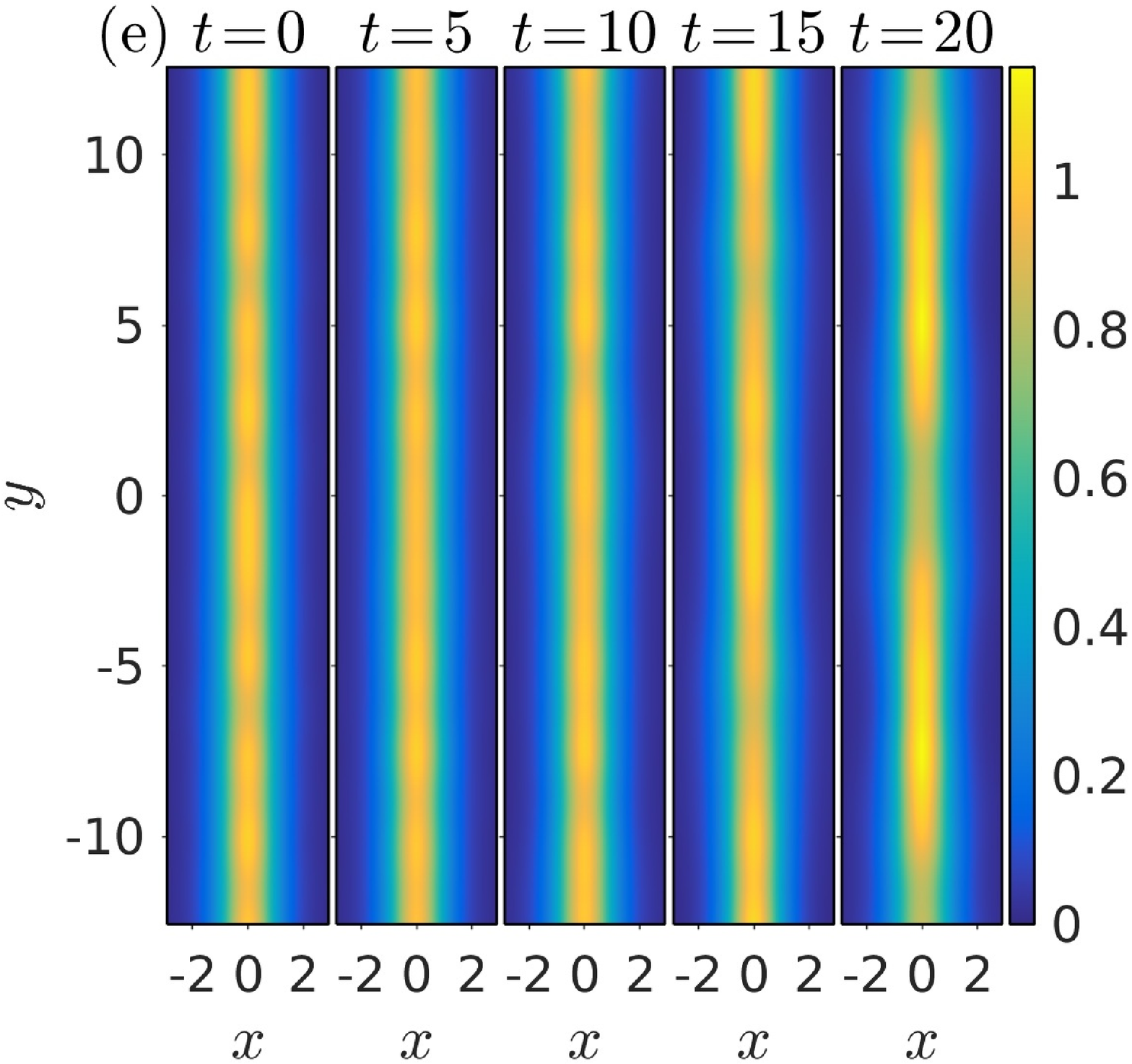}
\includegraphics[height=4.0cm]{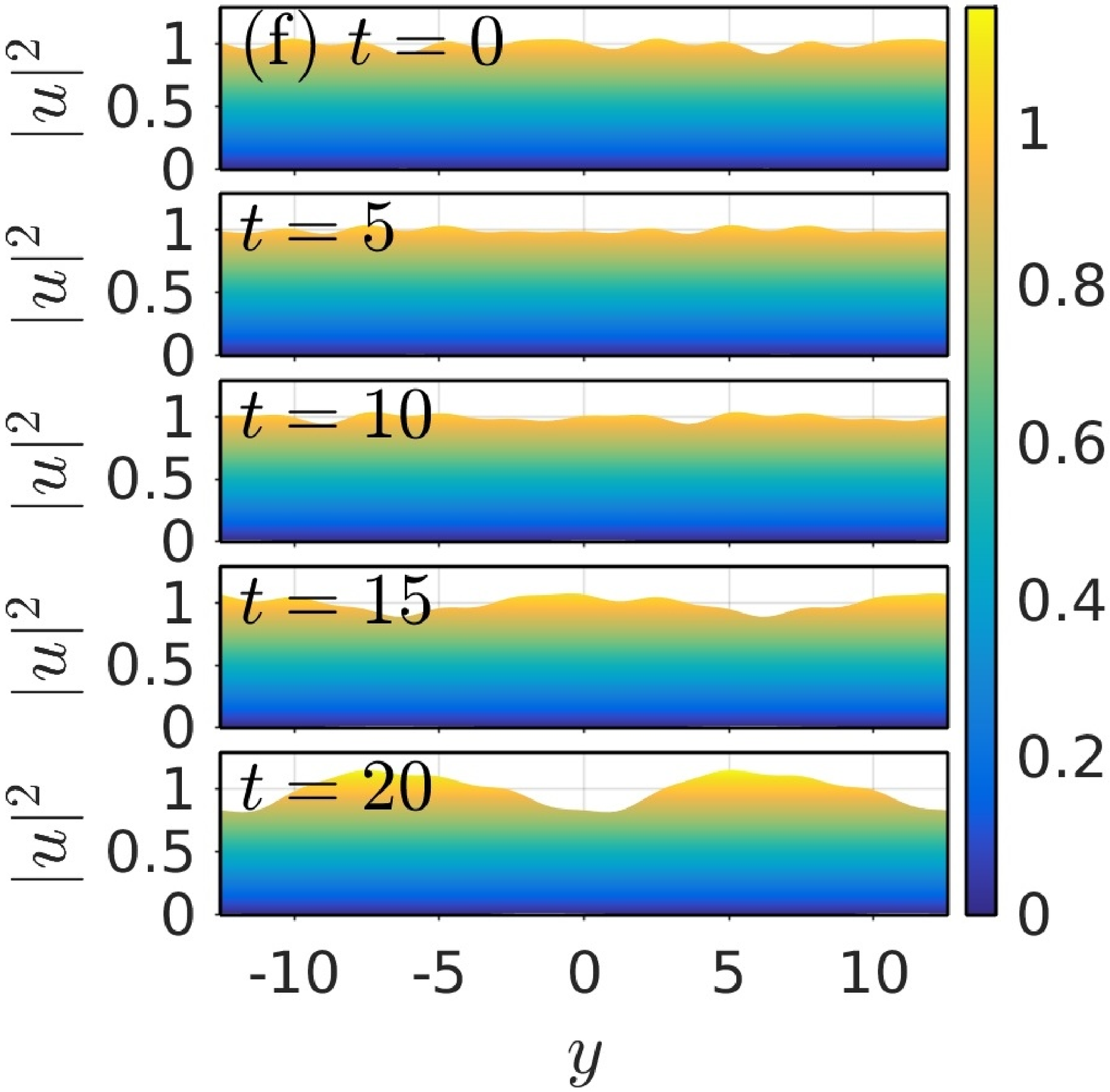}
\caption{(Color online) 
Influence of necking on the snaking dynamics. 
Panels (a) and (c) represent, respectively, top and side views of the 
evolution of the density of a bright-soliton stripe, to which solely 
a snaking perturbation is added initially, as $X_0=p(y)$
with the perturbation chosen as $p(y) = \sum_{j=1}^{5}\protect%
\varepsilon _{j}\,\sin (2\protect\pi jy/L_{y}+\protect\varphi _{j})$, where 
$\protect\varepsilon =0.01 $, $\protect\varphi _{j}=(j-1)L_{y}\protect\pi /10$.
The spatial domain is $[-L_{x},+L_{x}]\times
[-L_{y},+L_{y}]=[-20,+20]\times [-4\protect\pi ,+4\protect\pi ]$ (the panels
actually depict a zoom in the region $[-4,+4]\times [-L_{y},+L_{y}]$).
Panels (b) and (d) depict the same, but for a bright-soliton stripe
perturbed by \emph{both} snaking and necking perturbations.
In particular, we introduce a snaking perturbation by perturbing the position 
as above, $X_0=p(y)$, and we introduce a necking perturbation by 
perturbing the soliton height according to $A_0=1+p(y)$ where
$p(y)$ is the same perturbation defined above.
%
%
Panels (e) and (f) depict the effects of starting with only a necking
perturbation $A_0=1+p(y)$. Note that necking is unstable but that
snaking is not initiated.
}
\label{fig:dynamics1}
\end{figure}

To compare the different levels of the approximation ---corresponding to the
AI, full VA, and slowly varying VA---, we start with the stationary
soliton-stripe solution $u_{0}$ given in Eq.~(\ref{eq:BS}) at $t=0$ with
zero initial velocity ($v=0$) and centered at $x_{0}=0$ with $A=1$ [i.e.,
with $\mu =1/2$, as given by Eq.~(\ref{mu})] which we then perturb to create
the initial condition in the form of
\begin{equation}
u(x,y,t=0)=(A+\varepsilon_A A_0(y))\; u_{0}(x-\varepsilon_X X_0(y),y),  
\label{eq:BSpert}
\end{equation}
where $A_0(y)$ and $X_0(y)$ are, respectively, the amplitude and position
perturbations with amplitudes $\varepsilon_A$ and $\varepsilon_X$.
The perturbation may be 
(i) snaking applied to the initial position of the soliton stripe
along the $y$-direction (i.e., only $\varepsilon_X\not=0$), 
(ii) a necking deformation involving a perturbation of the 
soliton-stripe's amplitude in the $y$-direction. 
(i.e., only $\varepsilon_A\not=0$), or
(iii) both.
As corroboration of the VA results presented in 
Sec.~\ref{sec:VA}, we show in panels (a)--(d) of Fig.~\ref{fig:dynamics1} 
a typical example of the numerically simulated evolution involving a 
combination of snaking and necking perturbations. Panels (a) and (c) 
correspond to a bright-soliton stripe perturbed solely by a snaking 
perturbation, while panels (b) and (d) correspond 
to a mixture of snaking and necking. All the perturbations were
chosen with the same size and shape (see caption to Fig.~\ref{fig:dynamics1}
for details). The motivation for this set of simulations is to corroborate
the conclusion drawn from the VA, according to which the necking only
weakly affects snaking. Figure~\ref{fig:dynamics1} indeed suggests
that this is true. In this example, the snaking dynamics (see the top
panels) is almost identical for cases initiated with and without necking
perturbations.
We have also checked that other cases, corresponding to
variation of the relative amplitudes of the snaking and necking
perturbations, lead to similar results supporting this conclusion. 
On the other hand, we have observed that if the initial perturbation
is purely seeded in terms of necking (with no perturbation in snaking),
snaking is not initiated, at least for the time scales considered herein.
A typical example showing this is depicted
in panels (e) and (f) of Fig.~\ref{fig:dynamics1}.
%
This feature is quite 
useful as neither the AI or VA reduced model is able to appropriately capture 
the necking dynamics. Therefore, one may use the AI and VA methods to
exclusively follow snaking dynamics without the need to take care of the
necking perturbations. Accordingly, in what follows below we focus
exclusively on the snaking dynamics. Nonetheless, it is important to mention
that necking perturbations are always present (even if not introduced
initially), as snaking always induces the necking behavior, cf.~the example
depicted in the left panels of Fig.~\ref{fig:dynamics1}.

\begin{figure}[tbh]
\hspace{0.06cm} \includegraphics[width=1.0\columnwidth]{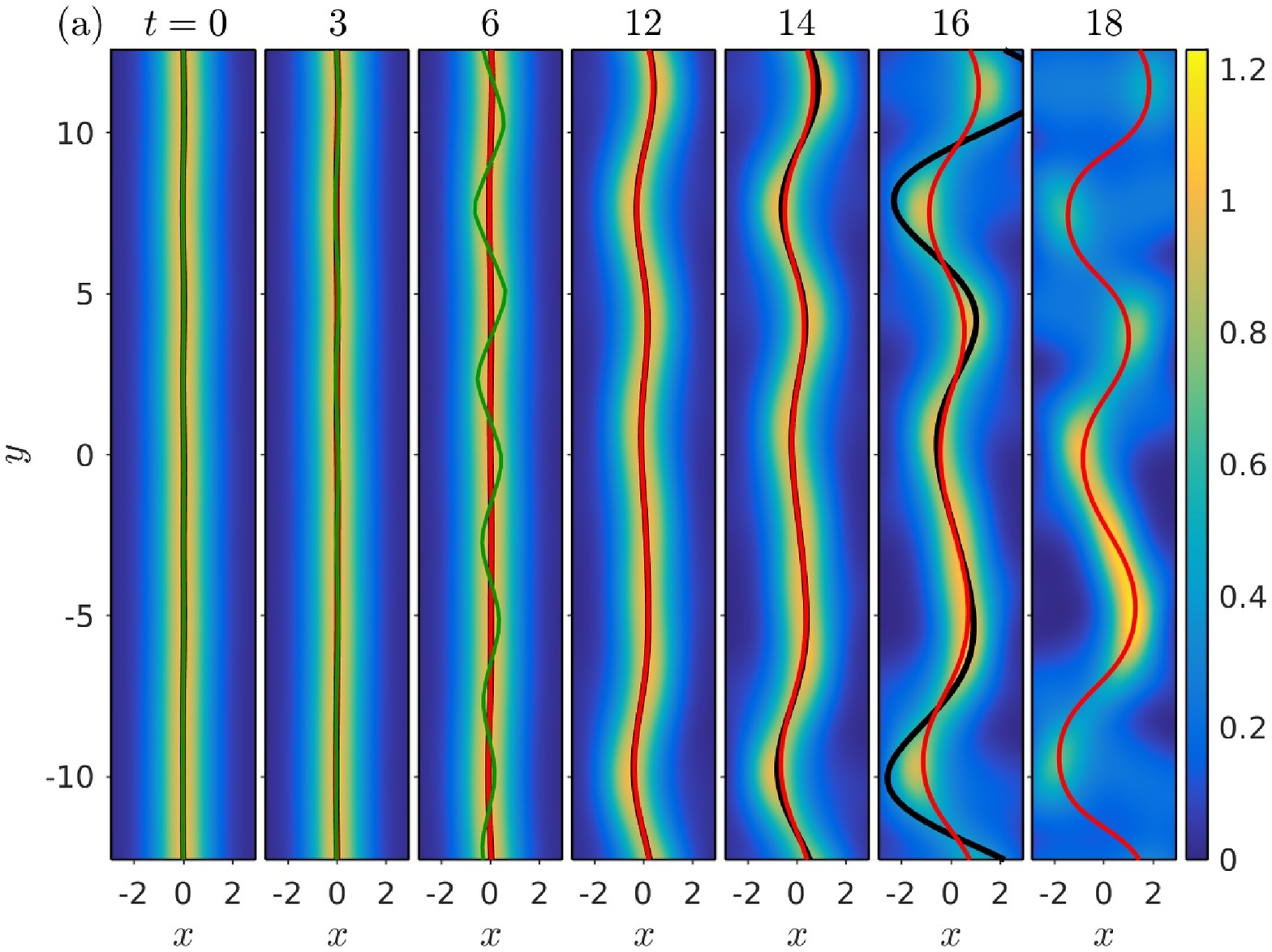}
\\[1.5ex]
\includegraphics[width=0.45\columnwidth]{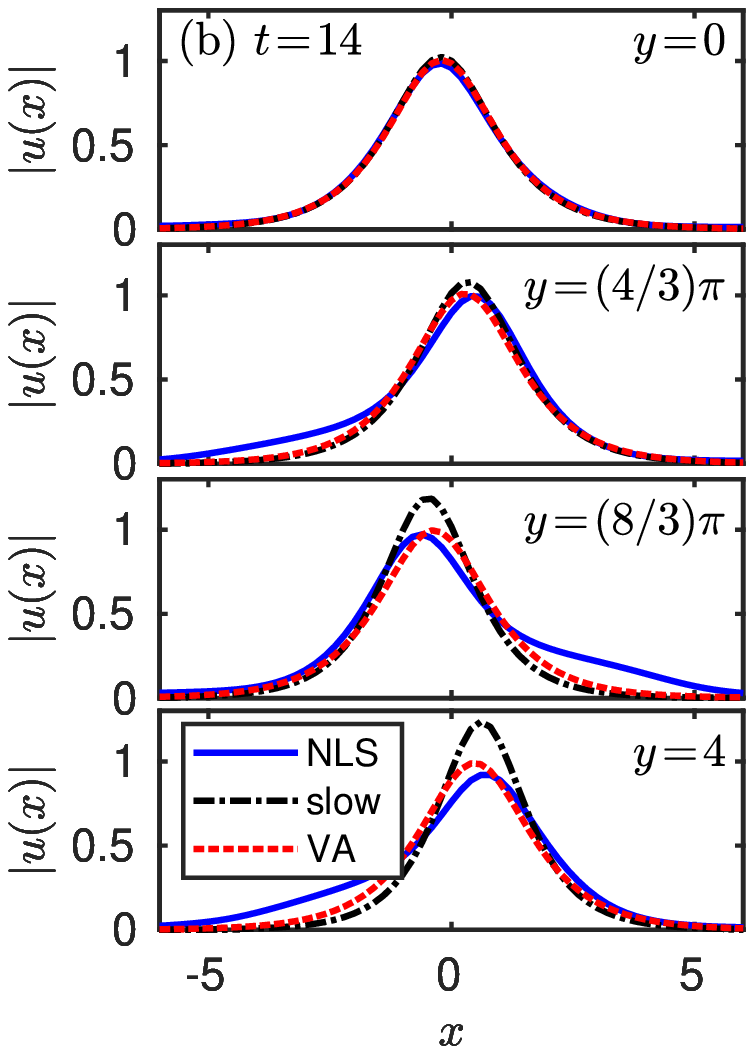} ~~
\includegraphics[width=0.45\columnwidth]{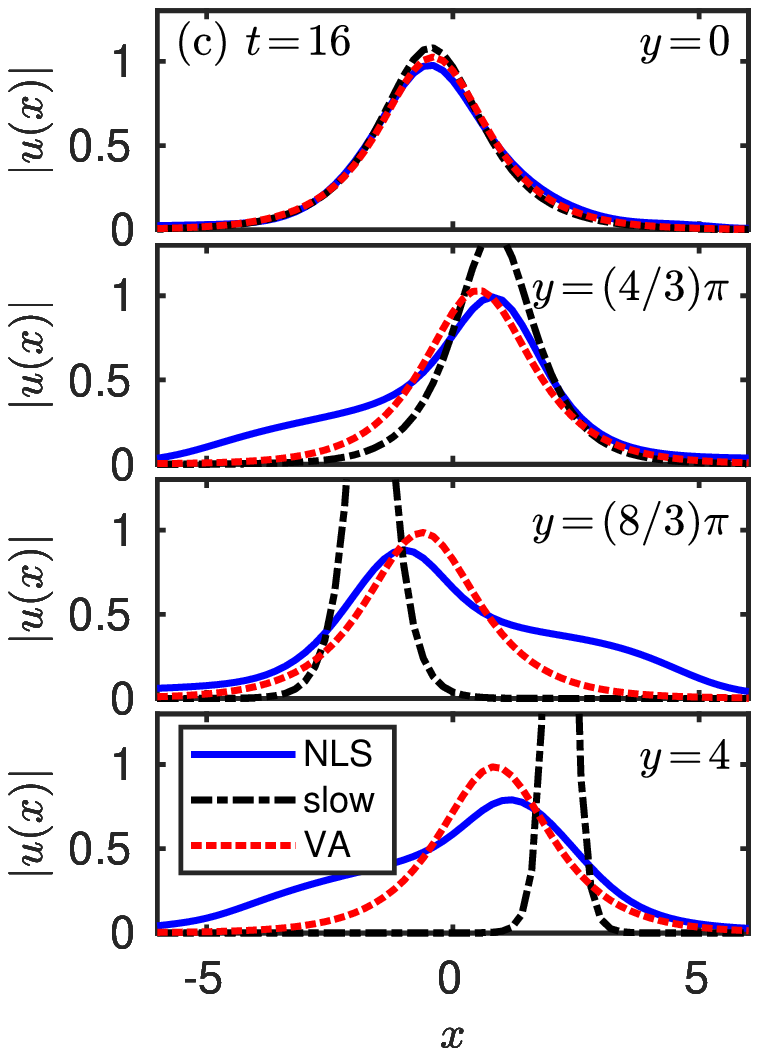}
\caption{(Color online) Comparison of the stripe's dynamics, as produced by
the simulated equation (\protect\ref{eq:NLS}), and by AI and VA reductions.
Panel (a) depicts the dynamics of a bright-soliton stripe initialized as in
case (a) of Fig.~\protect\ref{fig:dynamics1}, together with the reductions:
the AI, introduced as per Eq.~(\protect\ref{eq:AI}) (green curves), the full
VA, as per Eqs.~(\protect\ref{eq9a})--(\protect\ref{eq12}) (red curves), and
the slowly varying VA, defined by Eqs.~(\protect\ref{eq:A_slow})--(\protect
\ref{eq:theta_slow}) (black curves). Note that the AI and slowly varying VA
models blow up shortly after $t=6$ and $t=16$, respectively. Panels (b) and
(c) depict, at, respectively, $t=14$ and $t=16$, a few cross sections along 
$y=$ constant, produced by the full NLS simulations (blue solid curves),
together with the corresponding reconstructions of the stripe by means of
the VA (dashed red curves) and slowly varying VA (black dash-dotted curves).
The cross sections correspond, from top to bottom, to $y=0$, 
$y=4\protect\pi /3$, $y=8\protect\pi /3$, and $y=4\protect\pi =L_{y}$. }
\label{fig:dynamics2}
\end{figure}

We now compare the snaking dynamics obtained from simulations of 
Eq.~(\ref{eq:NLS}) and the AI and VA reductions. 
An example of this comparison is
depicted in Fig.~\ref{fig:dynamics2}. Panel (a) depicts the full simulations
for the snaking, along with \ results produced by the reductions, as
indicated in the caption. All models have the same linearization for long
wavelengths, therefore the dynamics for short times ($t<1$) is very similar
for both reductions, closely matching results of the direct simulations.
However, the AI reduction (shown by green curves) overestimates the growth
rate of all perturbations (cf.~Fig.~\ref{fig:spectra_AI_VA}) and thus its
predictions quickly cease to match results of the simulations. 
In fact, the
AI dynamics blows up shortly after $t=6$. On the other hand, the VA
predictions closely follow the simulated snaking at longer times. The slowly
varying VA (see black curves) is able to predict the snaking up to $t=14$.
However, shortly after $t=16$, this approximation blows up, as was the
case for the AI reduction. In contrast, the full VA method (see red curves)
is able to accurately follow the numerically simulated snaking at all times
up to the point were the stripe starts to break up into individual bright
patches ($t>14$). It is remarkable that even when the stripe is in the
process of breaking up (at $t\simeq 14$), up to the emergence of clearly
separated bright patches ($t>16$) the VA reduction still very closely
follows the location of the numerically simulated quite intricate
pattern of snaking.

To further confirm the validity of the VA, we depict in panels (b) and (c)
of Fig.~\ref{fig:dynamics2} several cross sections at $y=$ constant, as
produced by the full simulations of the NLS equation (\ref{eq:NLS}), and by
the VA reductions. At $t=14$ [panel (a)], the simulated NLS dynamics (solid
blue curves) starts to develop asymmetric tails that are more pronounced for
values of $y$ where the snaking is further away from $x=0$. Despite the fact
that the VA method produces, by construction, a symmetric
(i.e., even in $x$) profile, the VA
profiles closely follow the numerically simulated counterparts.
Notice the importance of the higher order terms in controlling both
the center position (which is still reasonably accurate for the VA),
but also importantly the amplitude of the solution; the absence of
these terms in the reduced VA can be clearly seen to lead to its
less accurate representation already at $t=14$.
However, at 
$t=16$ [panel (b)], the slowly varying VA starts to drift away from the
numerically produced NLS dynamics, in terms of snaking and necking alike,
the latter pertaining to the amplitude of the bright-soliton stripe. At this
stage, the amplitudes predicted by the slowly varying VA are several
times larger than the original one. Nonetheless,
even at this stage, when the numerically simulated $x$-profiles
are far away from the VA-based soliton (symmetric) ansatz (see, for
instance, the numerically simulated NLS profile at $t=16$ and $y=4\pi /3$),
the full VA is able to approximate well the location and amplitude of the stripe. In
addition to the typical example displayed in Fig.~\ref{fig:dynamics2}, we
have checked that other perturbations lead to similar results (not shown
here in detail). Thus, the direct simulations corroborate the validity of
the VA method especially in its full form,
and the limitations of the simplified AI reduction.

\section{Using the external potential to control and eliminate instabilities}
\label{sec:potential}

In this section, we aim to demonstrate how the external potential in 
Eq.~(\ref{eq:NLS}) can be used to attenuate and, possibly, eliminate
instabilities of the bright-soliton stripe. To suppress the snaking
instability, it is natural to propose a channel-shaped potential uniform in
the $y$-direction, with a transverse Gaussian profile:
\begin{equation}
V(x,y)=\frac{V_{0}}{\sqrt{2\pi }\sigma }\,e^{-\frac{x^{2}}{2\sigma ^{2}}},
\label{Vgauss}
\end{equation}
where $\sigma $ and $V_{0}<0$ are the width and strength of the Gaussian. 
$V_{0}>0$ (or, similarly, the edge of the 2D medium~\cite{surface}) may be
used to stabilize dark-soliton 
stripes~\cite{dark1,Adhikari,Ponz,Boaz,dark2,dark3}, while it
tends to enhance the instability of the bright one (not shown here in
detail).

\begin{figure}[tbh]
\includegraphics[width=\columnwidth]{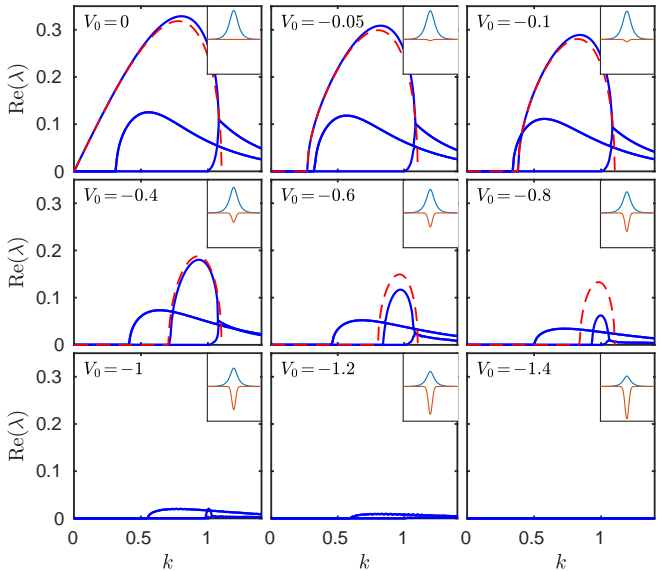}
\caption{(Color online) 
The use of the potential channel (\protect\ref{Vgauss}) to suppress the 
instabilities of the bright-soliton stripes.
Depicted is the real part of the instability spectra, following a variation of
the potential's strength $V_{0}$ for a constant width of potential 
$\sigma=1/2$. The insets depict a cross section, at
fixed $y$, of the stationary solution (the light blue upward-looking hump)
and the corresponding potential (the red downward hump), in the window of 
$[-7,+7]\times [-1.2,+1.1]$. As before, the stationary soliton stripe is
numerically computed by fixing $\protect\mu =1/2$ in Eq.~(\protect\ref{mu}).
Note the growth of the stabilizing effects as the potential gets deeper. The
snaking and necking instability branches are completely suppressed at 
$V_{0}\leq -1.2 $ and $V_{0}\leq -1.4$, respectively. Dashed red curves
depict the corresponding prediction for the linear stability, as provided by
VA, in the form of Eq.~(\protect\ref{eq:VA_disp}).}
\label{fig:spectra_sig0.5}
\end{figure}

Figure~\ref{fig:spectra_sig0.5} depicts the instability spectra, computed
and depicted as in Fig.~\ref{fig:SNmodesk0.5}, corresponding to the
stability of a stationary bright-soliton stripe with $\mu =1/2$ [see 
Eq.~(\ref{mu})], in the presence of potential (\ref{Vgauss}) with $\sigma =0.5$
and various values of $V_{0}$ (the insets show the respective transverse
profiles of the soliton and potential). As it seen in the figure, even a
very weak potential (see case of $V_{0}=-0.05$) shifts the
snaking-instability branch from the origin to the right and thus indicates
that long-wavelength perturbation cease being unstable. As the strength of
the potential channel increases, the snaking branch shifts further to the
right and, at the same time, it becomes shorter, showing not only that a
smaller interval of the snaking wavenumbers remains unstable, but also
that the
magnitude of the instability becomes smaller. It is worthy to note that the
necking-instability branch is also attenuated by the channel potential, in a
similar manner to the snaking branch. When the strength of the potential
attains the value $V_{0}=-1.2$ in Eq.~(\ref{Vgauss}), the snaking
instability is \emph{completely} \emph{eliminated}. The stabilization
becomes perfect at $V_{0}=-1.4$, when the necking instability is completely
suppressed too. We have checked that the randomly perturbed bright-soliton
stripe indeed develops no instability at $V_{0}\leq -1.4$ (not shown here in
detail). Thus, the appropriately designed channel potential is capable to
provide a {\em full stabilization} of the soliton stripes.

Some remarks are in order at this stage. Firstly, the surprising
stabilization of the necking ---while we were trying to suppress snaking---
might be understood by recalling that the necking-instability eigenmode in
the free space is not fully localized around the soliton's core, due to the
tails attached to it (in the $x$ direction). The channel potential is able
to curtail these tails, by trapping the soliton in the channel, thus helping 
to tame this eigenmode into an ``innocuous'' one.
On the other hand, we have tried other values of the channel's width $\sigma $ 
in Eq.~(\ref{Vgauss})], concluding that only values of $\sigma $ close to the
soliton's width provide complete suppression of the instabilities. In
particular, narrow potentials tend to produce a secondary instability,
whereby modes close to $k=0$ become unstable (not shown here in detail).
A detailed analysis of the secondary instability falls outside the scope of
the current work.

Finally, the prediction of the stabilization of the bright-soliton stripe
against the snaking perturbations in the channel potential, as produced by
the VA, has been tested too. Namely, the application of the linear-stability
condition, provided by Eq.~(\ref{eq:VA_disp}), to the Gaussian potential
given by Eq.~(\ref{Vgauss}), gives rise to the stability predictions
depicted by dashed red curves in Fig.~\ref{fig:spectra_sig0.5}. 
In the figure we only present the VA results for $-0.8 \leq V_0\leq 0$ since for
values of $V_0$ below $-0.5$ the VA prediction starts to deteriorate
and fails to give as accurate of a prediction for the snaking instability window. 
This drawback is a direct consequence of the choice
of reduced ansatz that we used in Eq.~(\ref{eq:BS}). In particular,
ansatz (\ref{eq:BS}) assumes that the height and inverse width of the 
soliton are {\em equal} (to $A$). We have checked that this assumption
approximately holds for all the statics and dynamics presented herein
in the absence of external potential ($V_0=0$). However, in the presence 
of the external potential ($V_0<0$), while the height of the bright soliton
does decrease as $V_0$ decreases, its width does not increase and
is kept approximately constant by the confining nature of the potential
(see insets in Fig.~\ref{fig:spectra_sig0.5}).
This shortcoming of the VA could be mended by considering an ansatz
including the width of the soliton as a new, independent, variational
parameter. However, this approach would also require the addition of a 
chirp (phase) variational parameter (as a conjugate variable to the 
soliton width) that would considerably complicate the VA methodology.
Such an extension falls outside of the scope of the present manuscript.
Nonetheless, it is important to stress that the results presented above
suggest that the VA is able to approximate the tendency
of the channel potential towards stabilization (i.e., towards
reduction of the snaking instability growth rate)
very well for moderate potential strengths, with 
$\left\vert V_{0}\right\vert <0.5$.


\section{Conclusions \& Future Challenges}

\label{sec:conclu}

We have studied the stability and dynamics of transverse instabilities,
mainly of the snaking type, for bright-soliton stripes in the 2D NLS
equations with the dispersion of the hyperbolic type. Such settings can be
readily implemented in terms of the spatiotemporal propagation in
self-focusing optical waveguides with the normal GVD (among other
physical settings). The analysis reported
here fully accounts for the mechanisms underlying the transverse snaking
instability of the stripe, with the help of approximate reductions of the
underlying NLS equation, based on the AI (adiabatic-invariant) and VA
(variational approximation) methods, which are applied to suitable
\textit{ans\"{a}tze}. While the applicability of the results provided by the
simplified AI approximation considered
here is limited, the VA method, applied to the soliton stripe
with parameters (position, velocity, amplitude, and phase) that depend not
only on time, but also on the longitudinal spatial coordinate, yields an
accurate description of the snaking stability for the stripe. Furthermore,
we have shown that the reduced dynamical equations of motion, which are also
produced by the VA, are quite efficient in describing the full nonlinear
dynamics of the unstable bright-soliton stripe, all the way to the stage at
which the stripe breaks up into individual bright spots. Further,
introducing an appropriately designed channel potential, we were able to
fully suppress the instabilities, of both the snaking and necking types. As
concerns the stabilization of the stripe against the snaking perturbations,
the VA provides good estimates of the (in)stability eigenvalues in the case
of small up to moderate strengths of the channel potential.

The present paper not only complements previous studies on the subject but
also puts forward efficient mechanisms for the control of the instabilities.
It is worth mentioning that the leading instability estimates, corresponding
to small wavenumbers of the snaking perturbations, originally due to
Zakharov and Rubenchik~\cite{Zakharov-Rubenchik:1974}, are captured and
improved by our VA method. We envisage potential applications of this
methodology to other higher-dimensional problems, such as dark-soliton
stripes, cf.~Refs.~\cite{dark1,Adhikari,Ponz,Boaz,dark2,dark3}. 
In the framework of the
present model, it may be relevant to construct stripes whose transverse
structure is similar not to the ground state of the Gaussian trapping
potential (\ref{Vgauss}), but to one of its excited bound states, provided
that the finite-depth potential supports such states (in terms of optics,
the latter condition corresponds to the condition that the channel waveguide
is a multimode one~\cite{multimode}). In fact, the self-trapping
nonlinearity may help to create bound states even if they do not exist in
the linear limit~\cite{Carr}. A challenging possibility is to develop the
analysis for bright-soliton \textit{filaments} in 3D media with the
hyperbolic dispersion, where the stabilization may be provided by a
fiber-like trapping channel burnt in the bulk material (an experimental
technique which can creates such channels is well known~\cite{Jena}).
Another significant challenge consists of improving the adiabatic
invariant methodology, by utilizing an amplitude dependent ansatz
in it and considering the resulting canonical formalism and the
associated equations of motion. This would be a significant
addition to the arsenal of methods of relevance to both bright
and dark filamentary structures. Such studies are currently
in progress and will be reported in future publications.

\section*{Acknowledgements}

LACA gratefully acknowledges the permit of absence from IPN and the
financial support of the Fulbright-Garc\'ia Robles program during his
research stay at SDSU. 
PGK and RCG gratefully acknowledge the support by the
National Science Foundation under grants PHY-1602994 and PHY-PHY-1603058.


\end{document}